\documentclass[structabstract]{aa}  
\usepackage{graphicx}
\usepackage{natbib}
\bibpunct{(}{)}{;}{a}{}{,}
\usepackage{txfonts}

\usepackage{scalefnt}
\usepackage{hyperref}
\usepackage{longtable,lscape}
\usepackage{xspace}
\usepackage{upgreek}
\usepackage{rotating}
\def\halpha{\hbox{H$\upalpha$}\xspace}
\def\brgamma{\hbox{Br$\upgamma$}\xspace}
\def\mum{\ensuremath{\upmu}m\xspace}
\def\Wbrgamma{\hbox{$W_\mathrm{Br\upgamma}$}\xspace}
\def\JH{\hbox{$J$$-$$H$}\xspace}
\def\HKs{\hbox{$H$$-$$K_\mathrm{s}$}\xspace}
\def\HK{\hbox{$H$$-$$K$}\xspace}
\def\KsLp{\hbox{$K_\mathrm{s}$$-$$L^\prime$}\xspace}
\def\EKsLp{\hbox{$E(K_\mathrm{s}$$-$$L^\prime)$}\xspace}

\def\J{\hbox{$J$}\xspace}
\def\H{\hbox{$H$}\xspace}
\def\K{\hbox{$K$}\xspace}
\def\Ks{\hbox{$K_\mathrm{s}$}\xspace}
\def\Lp{\hbox{$L^\prime$}\xspace}
\def\JHKs{\hbox{$JHK_\mathrm{s}$}\xspace}
\def\JHKsLp{\hbox{$JHK_\mathrm{s}L^\prime$}\xspace}
\def\HplusK{\hbox{$H\!+\!K$}\xspace}

\begin{document}
   \title{Protoplanetary disk evolution and stellar parameters\\of T\,Tauri binaries in Chamaeleon\,I\thanks{Based on observations collected at the European Organisation for Astronomical Research in the Southern Hemisphere, Chile. ESO Data ID: 086.C-0762.}}


   \author{S.~Daemgen\inst{1,2}\thanks{\email{daemgen@astro.utoronto.ca}}
          \and
          M.~G.~Petr-Gotzens\inst{1}
	  \and
	  S.~Correia\inst{3}
          \and
	  P.~S.~Teixeira\inst{4}
          \and
	  W.~Brandner\inst{5}
          \and
	  W.~Kley\inst{6}
          \and
	  H.~Zinnecker\inst{7,8}
          }

   \institute{European Southern Observatory, Karl-Schwarzschildstr. 2, 85748, Garching, Germany.
              \and
	      Department of Astronomy \& Astrophysics, University of Toronto, 50 St. George Street, Toronto, ON M5S 3H4, Canada
	      \and
	      Institute for Astronomy, University of Hawaii, 34 Ohia Ku Street, Pukalani, HI 96768, USA
	      \and
	      Institute for Astronomy, University of Vienna, T\"urkenschanzstrasse 17, A-1180 Vienna
	      \and
	      Max-Planck-Institut f\"ur Astronomie, K\"onigstuhl 17, 69117, Heidelberg, Germany
	      \and
	      Institut f\"ur Astronomie \& Astrophysik, Universit\"at T\"ubingen, Auf der Morgenstelle 10, 72076, T\"ubingen, Germany
	      \and
	      SOFIA Science Center, NASA-Ames Research Center, MS 232-12, Moffett Field, CA 94035, USA
	      \and
	      Deutsches SOFIA Institut, Univ.\ Stuttgart, Pfaffenwaldring 29, D-70569 Stuttgart, Germany
               }

   \date{}

 
  \abstract
   {}
   {This study aims to determine the impact of stellar binary companions on the lifetime and evolution of circumstellar disks in the Chamaeleon\,I (Cha\,I) star-forming region by measuring the frequency and strength of accretion and circumstellar dust signatures around the individual components of T\,Tauri binary stars.
}
   {We used high-angular resolution adaptive optics \JHKsLp-band photometry and 1.5--2.5\mum spectroscopy of 19 visual binary and 7 triple stars in Cha\,I -- including one newly discovered tertiary component -- with separations between $\sim$25 and $\sim$1000\,au. The data allowed us to infer stellar component masses and ages and, from the detection of near-infrared excess emission and the strength of Brackett-$\upgamma$ emission, the presence of ongoing accretion and hot circumstellar dust of the individual stellar component of each binary.}
   {Of all the stellar components in close binaries with separations of 25--100\,au, 10$^{+15}_{-5}$\% show signs of accretion. This is less than half of the accretor fraction found in wider binaries, which itself appears significantly reduced ($\sim$44\%) compared with previous measurements of single stars in Cha\,I. Hot dust was found around 50$^{+30}_{-15}$\% of the target components, a value that is indistinguishable from that of Cha\,I single stars. Only the closest binaries ($<$\,25\,au) were inferred to have a significantly reduced fraction ($\lesssim$\,25\%) of components that harbor hot dust.
Accretors were exclusively found in binary systems with unequal component masses M$_\mathrm{secondary}$/M$_\mathrm{primary}$\,$<$\,0.8, implying that the detected accelerated disk dispersal is a function of mass-ratio. This agrees with the finding that only one accreting secondary star was found, which is also the weakest accretor in the sample.}
   {The results imply that disk dispersal is more accelerated the stronger the dynamical disk truncation, i.e., the smaller the inferred radius of the disk. Nonetheless, the overall measured mass accretion rates appear to be independent of the cluster environment or the existence of stellar companions at any separation $\gtrsim$25\,au, because they agree well with observations from our previous binary study in the Orion Nebula cluster and with studies of single stars in these and other star-forming regions.}


   \maketitle
%

\section{Introduction}
Binary stars are among the most important branches of stellar formation. Owing to their observationally complex nature, the impact of a stellar companion on the evolution of the circumstellar disks, however, has yet to be explored in detail. Analytical and numerical studies consistently predict the outer truncation radius of a circumstellar disk to be less than one half of the binary separation \citep{art94,arm99,pic05}. Since typical binary separations are on the order of a few ten to a few 100\,au \citep{lei93,mat94}, disks around binary star components are significantly smaller than their 100--1000\,au single star counterparts -- a prediction qualitatively confirmed by the reduced flux in submm binary observations \citep{jen94,jen96,har12}. The resulting systems allow one to explore disk evolution with parameters that are distinct from single stars in terms of disk size, mass, and irradiation.
Observational exploration of these systems helps to advance the understanding of disk structure and evolution: the distribution of material in the disk, dust particle dynamics, and grain growth leave an imprint in the photometric and spectroscopic signatures of the star-disk systems. These can be compared with either predictions of disk evolution models \citep[e.g.,][]{kle08,zso11,mul12} or the distribution of planets in binary systems \citep{roe12}. Comparison with the latter constrains planet formation scenarios by reflecting whether the conditions in the modified disks, e.g., evolutionary timescales, are favorable for the formation of planets or not.

Results from previous studies of visual T\,Tauri binary stars draw an inconclusive picture: While the correlation of disk size and mass with time until the disk disperses is expected from evolutionary models \citep[e.g.,][]{mon07}, observations appear to return no clear signature. For example, mixed binary systems consisting of one component harboring a disk and one bare stellar component exist in both configurations -- the more massive or less massive component can be the disk-bearing component \citep[\emph{submitted}]{whi01,har03,pra03,mon07,dae12a,cor13}. Accordingly, the larger and more massive disk or smaller and less massive disk may disappear first. As in the single-star case, the individual binary components appear to lose their inner disk between $<$\,1\,Myr and $\sim$10\,Myr \citep{kra12}.

Nevertheless, binary disk lifetimes are typically shorter than in single stars and the less massive component loses its disk -- on average -- first \citep{mon07}. Most of the signatures, however, are blurred or may even remain undiscovered since the typically small sample sizes require combining data from different star-forming regions and from a wide mass range. This results in degeneracies when one tries to separate mutual influences in the parameter space, as stellar masses and the cluster environment are known to have significant influence on disk evolution. Larger sample sizes could so far only be investigated for spatially unresolved systems that measure, e.g., the total disk mass in a binary system. While \emph{binary} trends, suche as an accelerated disk dispersal, could be confirmed \citep{kra12,cie09}, variations with \emph{component} mass and age as well as differential component disk evolution were not accessible to these studies. 

Only few studies tackled these limitations by observing unbiased samples with coherent observational methods. Dedicated spatially resolved binary observations of the dense Orion Nebula cluster \citep[ONC;][]{dae12a,cor13} and the young Taurus association \citep{whi01} feature results which, however, require substantiation with data from other star-forming regions. On one hand, new observations can be used to decide which characteristics are universal and which are features of the individual regions under consideration. On the other hand, a careful selection of the target samples may allow one to extract information about disk properties as a function of crucial parameters such as, e.g., age and mass.

To expand the size of an available coherent dataset of disks in visual binaries, this paper focuses on the Chamaeleon\,I (Cha\,I) region. As part of the Chamaeleon complex, the Cha\,I dark cloud is close to the Sun \citep[160$\pm$15\,pc;][]{whi97} with an age distribution peaking between 2 and 3\,Myr \citep{luh07}. Of the $\sim$250 currently known stellar and substellar members $\sim$50\% show evidence of a circumstellar disk \citep{luh08b}. Close to 40 binaries and higher order multiples are known in a separation range between 16 and 1000\,au, which implies a multiplicity fraction of about 30\% -- higher by a factor of $\sim$1.9 than in the field \citep{laf08a}. Not only does Cha\,I's proximity grant access to spatially resolved visual binaries with separations as close as a few tens of au through using adaptive optics on 8\,m-class telescopes, but its older age compared with Taurus and the ONC can also provide first insights on disk evolution with time. 

The study presented in this paper explores a sample of 26 multiple stars in the Chamaeleon\,I star-forming region. The presence of accretion and/or hot dust in the circumstellar disks is evaluated for each individual stellar component using high-angular resolution near-infrared photometry and spectroscopy. This study complements our previous observations of the ONC \citep{dae12a,cor13} to explore the evolution of circumstellar disks in binaries as a function of age and environment. The observations are part of our ongoing efforts to characterize disk evolution as a function of stellar and binary parameters, with the goal of composing a coherent picture of star and planet formation in multiple stellar systems.

\section{Observations and data reduction}
\subsection{Sample definition}
Nineteen binaries and seven triple stars, including one newly discovered tertiary component to T\,33\,B, were observed in the Chamaeleon\,I star-forming region, with both SINFONI/VLT \HplusK-band integral field spectroscopy and close-in-time NACO/VLT \JHKsLp imaging between December 2010 and April 2011. Twenty-four of the targets were selected from the binary survey by \citet{laf08a} and two likely hosts of sub-stellar companions were additionally included (CHXR\,73, \citealt{luh06}; T\,14, \citealt{sch08}). This sample of 26 multiples is designed to contain nearly all multiples in Cha\,I in the separation range of 0\farcs2 to $\sim$6\arcsec\ known when the targets were selected, which is equal to $\sim$30--1000\,au at the distance of Cha\,I. The separation range was chosen to ensure high-quality spatially resolved observations of the individual binary components with ground-based adaptive optics-assisted spectroscopy. The brighter component of each binary was required to be brighter than $V\!=\!16.7$, $R\!=\!17$, and $K\!=\!13.5$ to serve as the adaptive optics guide star for our NACO and SINFONI observations. All multiples were selected to have at least one component with a spectral type (SpT) in the T\,Tauri range, i.e., G-type or later. While typically all components of each multiple fulfill this criterion, two binaries have primary stars of comparably early spectral type (T\,26\,A, G0; T\,41\,A, B9). These components are excluded from some of the discussion in Sect.~\ref{sec:ChaI:results} to minimize the impact of biases caused by their relatively high mass. Likewise, the two substellar companions to CHXR\,73 and T\,14 were excluded because they turned out to be too faint for most of the spectroscopic evaluation.

All multiples are confirmed members of Cha\,I, as inferred from their space velocity and indicators of youth \citep[position in H-R diagram, infrared-excess, lithium abundance, stellar activity,][]{luh04, luh08b}. Physical binarity was studied by \citet{laf08a} and \citet{vog12} through assessing the local stellar density and co-motion of binary components. While these authors found the probability of chance alignments to be $<$\,10$^{-4}$ for all but the widest binaries, physical association of the wide T\,6, T\,26, and T\,39 multiples was confirmed through multi-epoch imaging data. In addition, the spectroscopy from this study allows us to rule out background giants, and we did not detect significant changes in separation and position angle for any multiples of the sample when we compared with previously published measurements. The previously undetected companion to T\,33\,B is discussed and shown to be physically related in Sect.~\ref{sec:ChaI:T33}. In the following, we will treat all multiples as physically bound systems.

In every multiple, the brighter component in \Ks-band is referred to as the \emph{primary} or ``A'' component. This typically means that they are of earlier spectral type and thus higher mass than the other members of each system. In two cases (T\,39, Hn\,4) the inferred mass of the B component is higher than that of the brightest component in \Ks. T\,39\,B is a spectroscopic binary that we were unable to resolve \citep{ngu12}. Its mass is accordingly very likely overestimated and T\,39\,B not the most massive component of this higher-order multiple system. To be consistent with previous publications, we kept the identifiers as they are (the brighter component in \Ks is referred to as \emph{A}) but treated the more massive component as the primary for the derivation of, e.g., mass ratios.

\citet{ngu12} examined all but five of our targets for spectroscopic binarity and found CHXR\,28\,B, T\,31\,A, and T\,39\,B to show signs of spectroscopic binarity (see Table~\ref{tab:obs:ChaIobservations}). \citeauthor{ngu12} also reported CHXR\,47 and Hn\,4 as spectroscopic binary candidates. These authors found, however, that the known visual companions are likely identical to the spectroscopic companions in their data. Since it is not possible to measure separate spectral or photometric properties of the spectroscopic components from our observations, the combined properties were used where applicable and possible implications of the binary nature are highlighted. These targets are excluded from those parts of the discussion where knowledge of the individual component properties is required. 

Five of the six triple stars in this sample show a clear hierarchical structure. Whenever binary parameters (such as separation, mass ratios) are discussed, these triples are included as two binary stars, one consisting of the close binary system, the other one using the combined characteristics of the close binary (total system mass, center position, etc.) to compose a binary with the wide tertiary component. One triple, T\,39, does not show a distinct hierarchical structure and is either seen in a projection that veils an existing hierarchy, or its components are currently in an unstable non-hierarchical configuration (for a discussion see \citealt{laf08a}). We include this system as two binaries consisting of the AB and AC components, respectively (see Fig.~\ref{fig:obs:ChaIimaging} for the configuration).

A list of the targets, spectroscopic binarity, and the dates of the observations can be found in Table~\ref{tab:obs:ChaIobservations}.
\begin{sidewaystable*}
\vspace{0.75\textheight}
\caption[Cha~I targets and observations]{Cha\,I targets and observations}
\label{tab:obs:ChaIobservations}
\centering
\begin{tabular}{r@{\,\,}c@{\,\,}lr@{\,\,}c@{\,\,}lllccllccl}
\hline\hline
    \multicolumn{3}{c}{RA}&
    \multicolumn{3}{c}{DEC}&
    &
    &
    &
    pre-&
    \multicolumn{1}{c}{telluric}&
    \multicolumn{1}{c}{MJD}&
    \multicolumn{2}{c}{phot. ref.}&
    \multicolumn{1}{c}{MJD}\\
    \multicolumn{3}{c}{($^h$\,$^m$\,$^s$)}&
    \multicolumn{3}{c}{($^\circ$\,$^\prime$\,$^{\prime\prime}$)}&
    Identifier &
    Simbad Name\tablefootmark{a} &
    SB\tablefootmark{b} &
    optics\tablefootmark{c} &
    \multicolumn{1}{c}{reference} &
    \multicolumn{1}{c}{spectroscopy} &
    \multicolumn{1}{c}{$JHK_\mathrm{s}$} &
    \multicolumn{1}{c}{$L^\prime$} &
    \multicolumn{1}{c}{photometry}\\[0.5ex]
\hline
    10&55&59.759  &$-$77&24&40.07   & \object{T\,3}      &   Ass Cha T 2-3      &                  &  0.25   & Hip\,061193       & 55571.32071327  & S708-D\_9141  & HD\,75223  & 55610.19226973  \\
    10&58&16.774  &$-$77&17&17.06   & \object{T\,6}      &   Ass Cha T 2-6      & no               &  0.25   & Hip\,064501       & 55559.34191219  & S372-S\_9137  & HD\,77281  & 55602.28873422  \\
    11&01&18.750  &$-$76&27&02.54   & \object{CHXR\,9C}  &                      & no               &  0.1    & Hip\,061193       & 55571.34166417  & S372-S        & HD\,77281  & 55557.29178934  \\
      &  &        &     &  &        &                    &                      &                  &         & Hip\,050126       & 55587.20283114  &               &            &                 \\
    11&02&32.654  &$-$77&29&12.98   & \object{CHXR\,71}  &                      & no               &  0.1    & Hip\,050126       & 55542.34920563  & HD\,110621    & GL347A     & 55641.08828209  \\
      &  &        &     &  &        &                    &                      &                  &         & Hip\,033708       & 55565.35663229  &               &            &                 \\
    11&04&09.090  &$-$76&27&19.38   & \object{T\,14}     &   Ass Cha T 2-14     & no               &  0.25   & Hip\,060429       & 55590.21242836  & S708-D\_9141  & HD\,106965 & 55590.34649026  \\
    11&05&14.673  &$-$77&11&29.09   & \object{Hn\,4}     &                      & (SB2?)\tablefootmark{d}     &  0.025  & Hip\,040105       & 55592.29434212  & HD\,110621    & GL347A     & 55641.10665852  \\
    11&05&43.002  &$-$77&26&51.75   & \object{CHXR\,15}  &                      &                  &  0.1    & Hip\,050126       & 55587.21137326  & HD\,110621    & FS129      & 55653.12026904  \\
    11&06&28.774  &$-$77&37&33.16   & \object{CHXR\,73}  &                      &                  &  0.1    & Hip\,050126       & 55587.22603233  & HD\,110621    & FS129      & 55646.12072363  \\
      &  &        &     &  &        &                    &                      &                  &         & Hip\,067042       & 55588.29378704  &               &            &                 \\
    11&07&20.722  &$-$77&38&07.29   & \object{T\,26}     &   Ass Cha T 2-26     & no               &  0.25   & Hip\,060429       & 55590.15206476  & HD\,110621    & HD\,106965 & 55653.14097253  \\
    11&07&28.256  &$-$76&52&11.90   & \object{T\,27}     &   Ass Cha T 2-27     &                  &  0.1    & Hip\,067042       & 55678.05304754  & HD\,110621    & HD\,106965 & 55653.16332159  \\
    11&07&55.887  &$-$77&27&25.75   & \object{CHXR\,28}  &                      & CHXR28B=SB1?     &  0.1    & Hip\,064501       & 55594.23506763  & HD\,110621    & HD\,106965 & 55653.18627437  \\
    11&08&01.486  &$-$77&42&28.85   & \object{T\,31}     &   Ass Cha T 2-31     & T31A=SB2?        &  0.1    & Hip\,065181       & 55569.28001642  & HD\,110621    & HD\,106965 & 55653.20674135  \\
    11&08&02.975  &$-$77&38&42.59   & \object{ISO\,126}  &   ISO-ChaI 126       & no               &  0.1    & Hip\,040105       & 55589.31037568  & HD\,110621    & HD\,106965 & 55653.22875679  \\
    11&08&14.938  &$-$77&33&52.19   & \object{T\,33}     &   Ass Cha T 2-33     & no               &  0.25   & Hip\,060429       & 55590.16543515  & HD\,60778     & HD\,60778  & 55666.03397411  \\
    11&09&11.723  &$-$77&29&12.48   & \object{T\,39}     &   Ass Cha T 2-39     & T39B=SB1         &  0.25   & Hip\,060429       & 55590.20255577  & HD\,60778     & HD\,60778  & 55666.05782756  \\
    11&09&18.129  &$-$76&30&29.25   & \object{CHXR\,79}  &                      & no               &  0.1    & Hip\,061193       & 55589.34228160  & HD\,60778     & HD\,60778  & 55666.08169563  \\
    11&09&50.019  &$-$76&36&47.72   & \object{T\,41}     &   Ass Cha T 2-41     &                  &  0.1    & Hip\,067969       & 55589.26156993  & HD\,110621    & HD\,75223  & 55605.33691345  \\
    11&09&54.076  &$-$76&29&25.31   & \object{T\,43}     &   Ass Cha T 2-43     & no               &  0.1    & Hip\,040105       & 55589.33528044  & HD\,60778     & HD\,60778  & 55666.11141484  \\
    11&09&58.738  &$-$77&37&08.88   & \object{T\,45}     &   Ass Cha T 2-45     & no               &  0.1    & Hip\,067969       & 55589.27490141  & HD\,60778     & HD\,60778  & 55666.13248263  \\
    11&10&38.018  &$-$77&32&39.90   & \object{CHXR\,47}  &                      & (SB2)\tablefootmark{d}      &  0.025  & Hip\,040105       & 55592.28742058  & HD\,60778     & HD\,60778  & 55666.15952256  \\
    11&11&54.002  &$-$76&19&31.14   & \object{CHXR\,49NE}&                      &                  &  0.1    & Hip\,061193       & 55589.36918014  & HD\,110621    & HD\,75223  & 55606.24352530  \\
    11&12&24.415  &$-$76&37&06.41   & \object{T\,51}     &   Ass Cha T 2-51     & no               &  0.1    & Hip\,067969       & 55589.29062811  & HD\,60778     & HD\,60778  & 55670.07307583  \\
    11&12&42.689  &$-$77&22&23.05   & \object{T\,54}     &   Ass Cha T 2-54     & no               &  0.1    & Hip\,040105       & 55589.32408822  & S372-S\_9137  & HD\,77281  & 55602.31526911  \\  
    11&14&25.890  &$-$77&33&06.10   & \object{Hn\,21}    &                      & no               &  0.25   & Hip\,040105       & 55592.32103162  & HD\,110621    & HD\,75223  & 55606.29302649  \\
    11&14&50.319  &$-$77&33&39.04   & \object{B\,53}     &   BYB 53             & no               &  0.1    & Hip\,040105       & 55589.30224627  & HD\,110621    & HD\,75223  & 55606.34424266  \\
    11&18&20.600  &$-$76&21&58.00   & \object{CHXR\,68}  &                      & no               &  0.25   & Hip\,060429       & 55590.19058466  & HD\,60778     & HD\,60778  & 55666.18125080  \\
\hline
\end{tabular}
   \vspace{-4ex}\begin{minipage}{1.\textwidth}\tablefoot{
      \tablefoottext{a}Identifier to be used with Simbad (\url{http://simbad.u-strasbg.fr/simbad/}) if different from identifier in the 3rd column.
      \tablefoottext{b}Spectroscopic binarity reference: \citet{ngu12}
      \tablefoottext{c}SINFONI pre-optics used (see text for definition)
      \tablefoottext{d}The spectroscopic companions to Hn\,4 and CHXR\,47 are probably identical to the visual companions in this paper \citep[see][]{ngu12}.
   }\end{minipage}
\end{sidewaystable*}

\subsection{NACO imaging: photometry and astrometry}
\subsubsection{Observations and data reduction}
All targets were observed with NAOS-CONICA \citep[NACO;][]{len03,rou03} imaging on UT4 of the Very Large Telescope (VLT). The adaptive optics system NAOS always used the brightest multiple component (or the binary itself for close systems) as a guide star. All components of each target were observed in the same field of view in four broad-band filters (\JHKsLp) during a single night to minimize the impact of variability. Observations in \JHKs use the N90C10 dichroic and the S13 camera with a pixel scale of 13.26\,mas/pix, \Lp uses the IR wavefront sensor and the L27 camera with 27.1\,mas/pix. Typical values of measured FWHM of our point sources are 0\farcs074 in \J, 0\farcs070 in \H,  0\farcs077 in \Ks, and 0\farcs11 in \Lp. The derotator was used to keep the north/south direction aligned with the columns of the detector. All targets were observed in a five-position on-source dither pattern to allow for sky subtraction. Total integration times per target and filter are between 50 and 800\,sec in \JHKs-band and equal to 176\,sec in \Lp, with the exception of the substellar companion host CHXR\,73 of 528\,sec. Photometric standard stars were always observed during the same night with the same instrumental setup as the science observations as part of the standard NACO calibration plan.

Pipeline-reduced data products were provided in the ESO data package for all science targets. A subsample was compared to images manually reduced with the \emph{NACO pipeline} in \emph{Gasgano}\footnote{v2.4.0, \url{http://www.eso.org/sci/software/gasgano/}} with the recommended settings to confirm their usability for photometric and astrometric analysis. The difference between the two reductions was negligible (photometry, noise), thus the provided data products were used for the photometry. Manual pipeline reduction, however, was required for the photometric standards, since no reduced data were available. For both the standards and science targets, the pipeline performs flat-fielding using lamp-flats and sky subtraction. After all the dithered images per target and filter were shifted and averaged, residual bad pixels were removed by substitution with the median value of the surrounding good pixels.

\begin{figure*}
  \centering
  \includegraphics[angle=0,height=0.95\textheight]{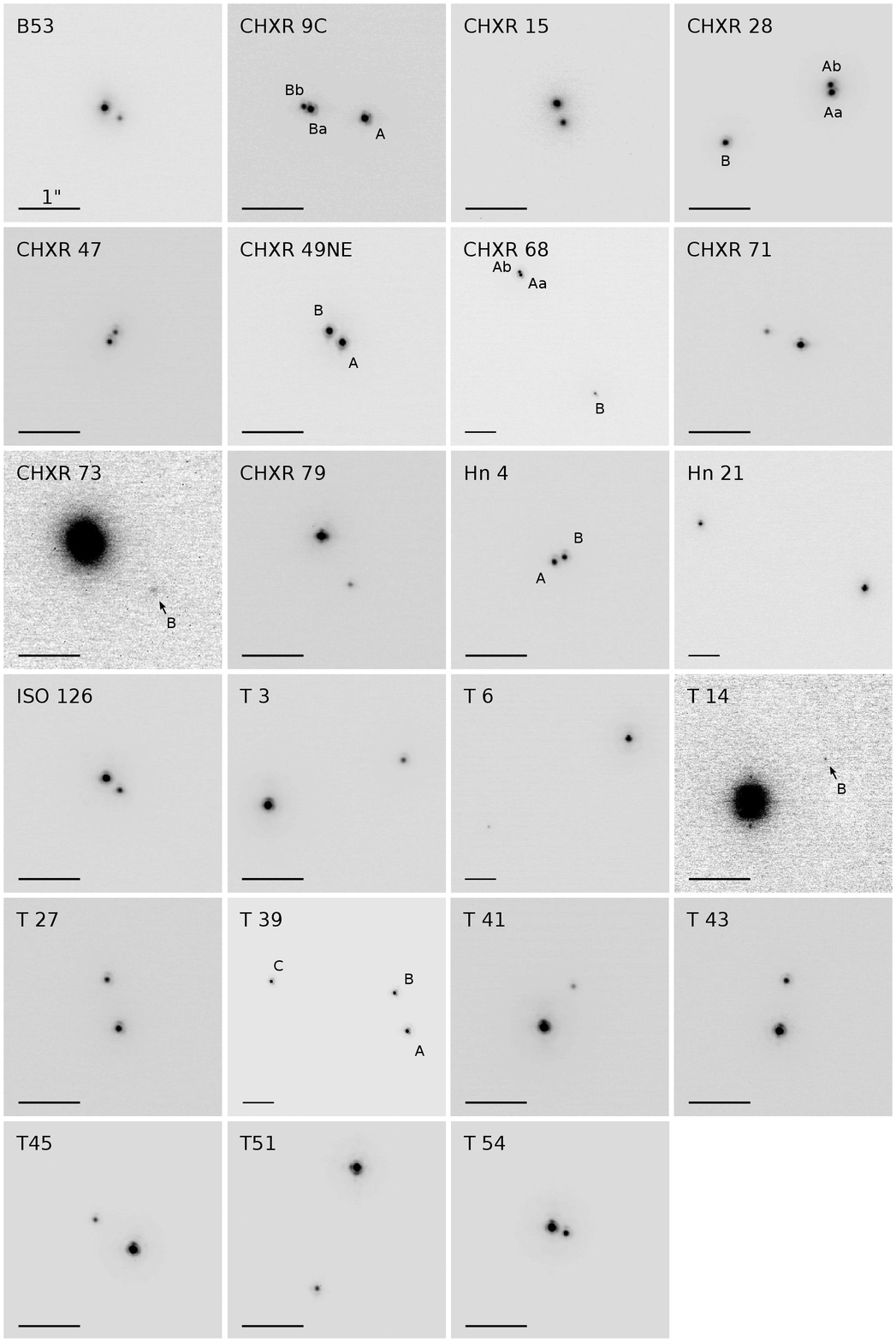}
  \caption[NACO Ks-band imaging of Cha~I binaries]{NACO \Ks-band imaging of all target binaries and triples (except T\,26, T\,31, and T\,33 in Fig.~\ref{fig:obs:ChaIT26T31}). Components are marked with their component identifier when it is not clear whether they are primary or secondary. The bar in each subpanel has a length of 1\arcsec. The color stretch is linear. North is up and east to the left.}
  \label{fig:obs:ChaIimaging}
\end{figure*}

\begin{figure*}
  \centering
  \includegraphics[angle=0,width=0.32\textwidth]{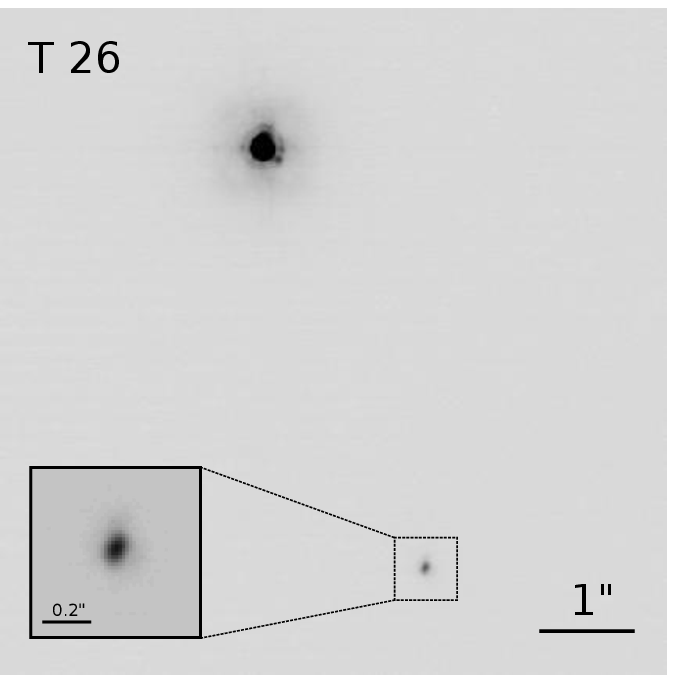}\hfill
  \includegraphics[angle=0,width=0.32\textwidth]{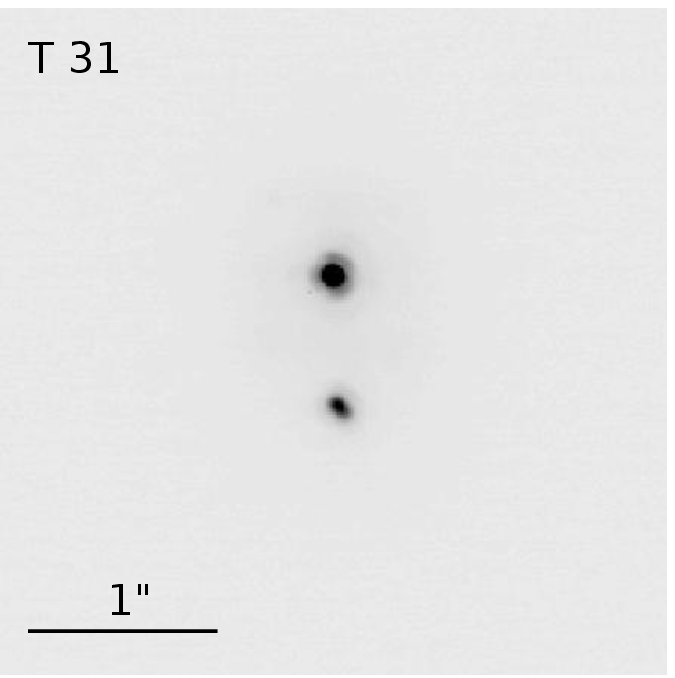}\hfill
  \includegraphics[angle=0,width=0.32\textwidth]{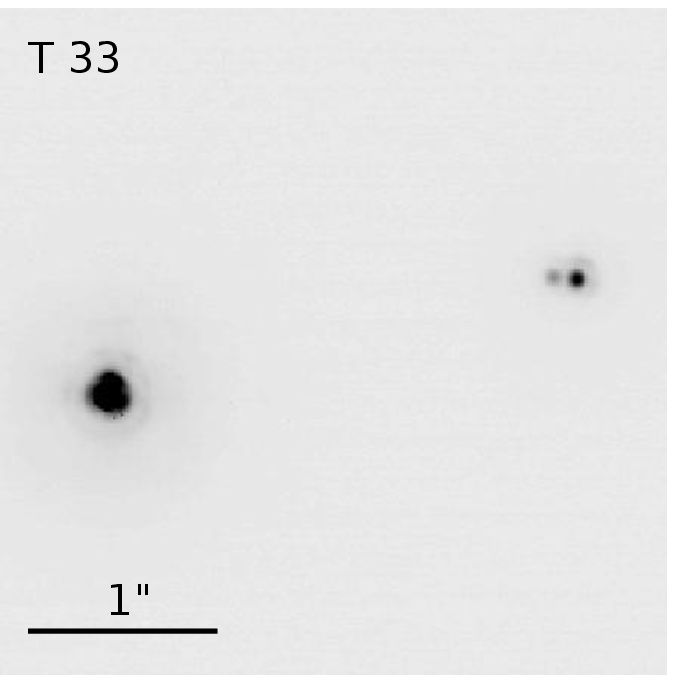}
  \caption{Close binary components of T\,26 (left) and T\,31 (middle), i.e., the outhern component in both panels, could not be spatially resolved with NACO \Ks-band imaging despite the good FWHM of $\sim$0\farcs070. From the apparent elongation we estimate that the separation of T\,26\,BaBb is not much larger than 0\farcs05 and just at the FWHM for T\,31\,BaBb. The \Ks-band image of T\,33 (right) shows the newly discovered faint tertiary component close to T\,33\,B. The physical association of all three components is discussed in Sect.~\ref{sec:ChaI:T33}. Image stretches and orientation are the same as in Fig.~\ref{fig:obs:ChaIimaging}.}
  \label{fig:obs:ChaIT26T31}
\end{figure*}

Images of all spatially resolved previously known multiples in the \Ks-band are shown in Fig.~\ref{fig:obs:ChaIimaging}. 
The close binary companions in the triple systems T\,26 and T\,31 could not be resolved. Their binarity is revealed only by their elongated shapes (Fig.~\ref{fig:obs:ChaIT26T31}) in these observations with a FWHM of $\sim$0\farcs07.
The estimated separations are $\lesssim$\,0\farcs05 (T\,26B) and $\sim$0\farcs07 (T\,31B). The tertiary component to T\,33 has not been reported before (Fig.~\ref{fig:obs:ChaIT26T31}) and its physical association with the system is assessed in Sect.~\ref{sec:ChaI:T33}.

\subsubsection{Photometry and astrometry}\label{sec:obs:ChaIphotometry}
Using a combination of calibrated and differential component photometry, \JHKsLp photometry of all target components was measured.

Differential photometry was obtained through simultaneous aperture photometry of pairs of stars in the image (typically the two binary components). Triple stars allowed using accurate PSF photometry with the {\tt daophot} task in \emph{IRAF} using the PSF of the wide tertiary component as a reference. Despite possible isoplanatic effects of the adaptive optics correction, subtracting of the flux-scaled reference resulted in residuals with typical amplitudes $\ll$1\% of the original component flux.

Separations and position angles, were inferred from the results obtained with \emph{IRAF} {\tt daofind}, which returns stellar positions with an accuracy of a fraction of a pixel, or $\lesssim$0\farcs01. The derotator was used to keep north up, so the inferred position angles are assumed to be absolute with a systematic uncertainty of 0.5$^\circ$.

The photometric calibration was performed with photometric standard stars observed during the same night or a night as close in time as possible, where the latter case is validated by the stablility of the zeropoint over the duration of several nights (rms error$\sim$0.02\,mag). Apertures of $\sim$0\farcs5--1\farcs9 radius were used, either enclosing one binary component or, for close companions, both components of a binary. We extracted the standard star photometry with the same individual apertures to avoid aperture corrections. Errors of the derived zeropoints caused by the variable adaptive optics PSF wings were estimated to be $<$\,0.05\,mag. Apparent magnitudes were then calculated considering normalized instrumental magnitudes, airmasses, Paranal extinction coefficients, and photometric zeropoints. The resulting photometry is listed in Table~\ref{tab:obs:ChaIphotometry} together with the measured separations and position angles.

Two stars in the sample, T\,26\,A and T\,33\,A, turned out to be too bright for observations in \Lp-band with NACO. Their saturated \Lp photometry cannot be used for further evaluation. The brown dwarf companion candidate T\,14\,B remains undetected in \J.

All \JHKs photometry presented in this section was observed in or converted to the 2MASS system \citep{cut03}, \Lp photometry was converted to the ESO system when necessary \citep{van96}. Photometry with NACO needs no transformation to be compatible with this system and the photometry of the standard stars is available in these filters\footnote{\url{http://www.eso.org/sci/facilities/paranal/}\\\url{instruments/naco/tools/imaging_standards/naco_lw_lp.html}}. For all other necessary transformations of \JHKsLp photometry (e.g. CTTS locus, extinction vector; Sect.~\ref{sec:ChaI:colorcolor}) the recipes of \citet{bes88} and \citet{car01} were used.

Some targets were observed multiple times for various reasons like bad weather, poor Strehl correction, or a missing photometric reference in one or more filters. The subset of repetitions that returned data of good quality in one or more filters were reduced and measured as described. More than half of all targets with repeated measurements exhibited significant photometric variability of up to $\sim$0.2\,mag and differential magnitude differences of $\lesssim$0.1\,mag. A scatter of this magnitude can be caused by several mechanisms, including intrinsic variability through chromospheric activity, variable accretion, or occultation from a warped inner disk as a result of magnetospheric star-disk interaction \citep[e.g.,][]{car01a,bou99,ale10}.

Despite occasional re-observations in individual filters, for all targets only one \emph{complete} set of observations exists, i.e., all of \JHKsLp were observed during one night together with photometric references. These observations were used for subsequent evaluation.

\subsection{SINFONI H+K-band IFU spectroscopy\label{sec:ChaI:obs}}
\subsubsection{Observations and data reduction}
Integral field spectroscopy was obtained with SINFONI \citep{eis03,bon04} on the VLT. All observations used the adaptive optics module, locked on the brightest multiple component or close binary. The offered \HplusK configuration was used to obtain simultaneous spectroscopy from $\sim$1.5 to $\sim$2.5\,\mum for an array of 32$\times$64 spatial pixels. Depending on the separation of the target binary, one of the 0.25, 0.1, or 0.025 pre-optics was used, resulting in a final spatial pixel scale in the reduced images (after resampling) of 0\farcs125/pix, 0\farcs05/pix, and 0\farcs0125/pix. The spectral resolution is $R\!\approx\!1500$ with $\sim$5\,\AA/pix in dispersion direction. The typical FWHM is $\sim$0\farcs10--0\farcs25, sufficient to spatially resolve all binaries and wide tertiary components in the sample. All close components in triples, however, are only marginally resolved and only the combined spectra were used for further evaluation. Except for CHXR\,73, all targets were observed in a four-position dither pattern. In the two larger scales (0.25 and 0.1 pre-optics) all four dithers were on source; the two closest binaries (Hn\,4 and CHXR\,47) were observed with alternating on- and off-source dithers with the 0.025 pre-optics. The total integration time was 2\,min per target. To increase the S/N of the faint secondary around CHXR\,73, this target was observed with 16 dither positions totaling 32\,min of integration time. Each science observation was accompanied by the observation of a bright telluric standard star of known spectral type (B2 to B6) with identical system settings at similar airmass. 

\begin{table*}
{\scalefont{0.93}
\caption[Individual component apparent magnitudes]{Individual component apparent magnitudes}
\label{tab:obs:ChaIphotometry}
\centering
\begin{tabular}{lccccr@{$\,\pm\,$}lr@{$\,\pm\,$}lr@{$\,\pm\,$}lr@{$\,\pm\,$}lr@{$\,\pm\,$}lc}
\hline\hline
      & 
      \multicolumn{2}{c}{separation\tablefootmark{a}} &      
      PA\tablefootmark{a} & 
      &
      \multicolumn{2}{c}{$J$\tablefootmark{b}} &
      \multicolumn{2}{c}{$H$\tablefootmark{b}} &
      \multicolumn{2}{c}{$K_\mathrm{s}$\tablefootmark{b}} &
      \multicolumn{2}{c}{$L^\prime$\tablefootmark{b}} &
      \multicolumn{2}{c}{$A_V^\mathrm{phot}$} &
      \multicolumn{1}{c}{$E_{K_\mathrm{s}-L^\prime}$}\\[0.5ex]
      \multicolumn{1}{l}{Name} &
      [arcsec]& 
      [au]& 
      [$^\circ$]&      
      &
      \multicolumn{2}{c}{[mag]} &
      \multicolumn{2}{c}{[mag]} &
      \multicolumn{2}{c}{[mag]} &
      \multicolumn{2}{c}{[mag]} &
      \multicolumn{2}{c}{[mag]} &
      \multicolumn{1}{c}{[mag]}\\[0.5ex]
      \hline
B\,53		& 0.28 &  45 & 235.8 & A &    10.80&0.03           &    10.03&0.04    &     9.87&0.02    &     9.66&0.01      &      0.4&0.5            &      0.03  \\            
		&      &     &       & B &    11.96&0.05           &    11.44&0.06    &    11.30&0.04    &    10.83&0.03      &      0.0&0.5            &      0.21  \\[0.5ex]     
CHXR\,9C	& 0.90 & 144 &  79.7 & A &    10.83&0.04           &    10.01&0.03    &     9.84&0.06    &     9.76&0.02      &      0.7&0.6            &   $-$0.11  \\            
		& 0.12 &  19 &  69.5 & Ba&    11.15&0.04           &    10.33&0.03    &    10.16&0.06    &    10.08&0.02      &      0.7&0.6            &   $-$0.12  \\            
                &      &     &       & Bb&    11.82&0.04           &    11.06&0.03    &    10.84&0.06    &    10.65&0.02      &      0.5&0.6            &      0.00  \\[0.5ex]     
CHXR\,15	& 0.31 &  50 & 198.2 & A &    11.91&0.03           &    11.30&0.04    &    10.99&0.02    &    10.40&0.04      &      1.0&0.3            &      0.20  \\            
		&      &     &       & B &    12.38&0.04           &    11.80&0.04    &    11.53&0.02    &    11.02&0.05      &      1.1&0.4            &      0.12  \\[0.5ex]     
CHXR\,28    	& 1.84 & 294 & 117.1 & Aa&     9.99&0.03           &     9.03&0.04    &     8.77&0.02    &     8.62&0.02      &      2.0&0.5            &   $-$0.05  \\            
		& 0.12 &  20 &   7.6 & Ab&    10.44&0.03           &     9.38&0.04    &     9.09&0.02    &     8.81&0.02      &      2.7&0.5            &      0.04  \\            
		&      &     &       & B &    10.46&0.03           &     9.44&0.04    &     9.15&0.02    &     9.00&0.02      &      2.5&0.5            &      0.17  \\[0.5ex]     
CHXR\,47	& 0.17 &  27 & 328.9 & A &    10.34&0.03           &     9.42&0.03    &     8.87&0.03    &     8.27&0.03      &      3.6&0.2            &      0.32  \\            
		&      &     &       & B &    10.80&0.04           &     9.83&0.03    &     9.35&0.04    &     8.77&0.03      &      3.6&0.3            &      0.29  \\[0.5ex]     
CHXR\,49NE	& 0.27 &  43 &  48.6 & A &    10.99&0.03           &    10.22&0.04    &    10.02&0.02    &     9.64&0.01      &      1.4&0.3            &      0.12  \\            
		&      &     &       & B &    11.09&0.03           &    10.37&0.04    &    10.15&0.02    &     9.73&0.01      &      0.9&0.3            &      0.14  \\[0.5ex]     
CHXR\,68    	& 4.40 & 704 & 212.1 & Aa&    10.59&0.03           &     9.89&0.04    &     9.61&0.02    &     9.67&0.04      &      0.4&0.5            &   $-$0.22  \\            
                & 0.09 &  15 &  21.9 & Ab&    10.74&0.03           &    10.07&0.04    &     9.84&0.02    &     9.71&0.04      &      0.0&0.5            &   $-$0.01  \\            
		&      &     &       & B &    11.41&0.03           &    10.67&0.03    &    10.40&0.03    &    10.28&0.03      &      0.6&0.4            &   $-$0.06  \\[0.5ex]     
CHXR\,71	& 0.56 &  90 &  68.5 & A &    11.55&0.03           &    10.69&0.04    &    10.45&0.02    &    10.24&0.02      &      1.2&0.5            &   $-$0.02  \\            
		&      &     &       & B &    13.12&0.03           &    12.41&0.04    &    12.06&0.02    &    11.27&0.03      &      0.0&0.3            &      0.53  \\[0.5ex]     
CHXR\,73	& 1.29 & 206 & 234.6 & A &    12.64&0.35           &    11.40&0.04    &    10.94&0.02    &\multicolumn{2}{c}{(\ldots)\tablefootmark{c}} &  4.5&2.5 & \ldots   \\ 
		&      &     &       & B &    17.84&0.37           &    16.35&0.10    &    15.47&0.16    &\multicolumn{2}{c}{\ldots}&      6.7&3.1            &     \ldots  \\[0.5ex]     
CHXR\,79	& 0.88 & 140 & 211.0 & A &    11.84&0.04           &    10.31&0.03    &     9.32&0.03    &     8.03&0.03      &      6.9&0.3            &      0.76  \\            
		&      &     &       & B &    13.84&0.05           &    12.50&0.04    &    11.80&0.04    &    10.87&0.08      &      5.6&0.4            &      0.38  \\[0.5ex]     
Hn\,4		& 0.17 &  27 & 295.2 & A &    11.65&0.03           &    10.76&0.04    &    10.45&0.02    &    10.17&0.01      &      1.7&0.5            &   $-$0.07  \\            
		&      &     &       & B &    11.70&0.03           &    10.77&0.04    &    10.47&0.02    &    10.32&0.01      &      1.9&0.5            &   $-$0.17  \\[0.5ex]     
Hn\,21 	        & 5.48 & 876 &  68.5 & A\tablefootmark{d} & 12.09&0.03 &11.14&0.04    &    10.70&0.02    &    10.18&0.02      &      3.5&0.3            &      0.10  \\            
	        &      &     &       & B\tablefootmark{d} & 12.90&0.04 &12.09&0.04    &    11.56&0.03    &    10.77&0.03      &      1.5&0.3            &      0.38  \\[0.5ex]     
ISO\,126	& 0.28 &  45 & 228.8 & A &    12.55&0.03           &    10.48&0.04    &     8.87&0.02    &     6.83&0.02      &     10.5&0.3            &      1.87  \\            
		&      &     &       & B &    12.70&0.03           &    11.21&0.05    &    10.11&0.04    &     8.33&0.02      &      4.3&0.2            &      1.41  \\[0.5ex]     
T\,3		& 2.21 & 354 & 288.4 & A &    10.97&0.03           &     9.64&0.02    &     8.58&0.03    &     7.12&0.01      &      4.9&0.2            &      1.05  \\            
		&      &     &       & B &    11.62&0.03           &    10.86&0.02    &    10.33&0.03    &     9.55&0.02      &      1.2&0.5            &      0.50  \\[0.5ex]     
T\,6		& 5.12 & 819 & 122.3 & A &     8.76&0.04           &     8.23&0.03    &     7.70&0.05    &     6.90&0.01      &      0.0&0.5            &      0.76  \\            
		&      &     &       & B &    12.00&0.08           &    11.65&0.06    &    10.95&0.08    &    10.39&0.02      &      0.8&0.3            &      0.26  \\[0.5ex]     
T\,14		& 2.67 & 427 & 299.6 & A &     9.45&0.02           &     8.88&0.02    &     8.48&0.03    &     7.66&0.01      &      0.0&0.5            &      0.72  \\            
		&      &     &       & B &\multicolumn{2}{c}{\ldots}&    15.01&0.36    &    16.54&4.16    &    13.01&0.15      &\multicolumn{2}{c}{\ldots}&     \ldots  \\[0.5ex]     
T\,26		& 4.57 & 731 & 201.3 & A &     7.89&0.03           &     7.06&0.04    &     6.45&0.02    &\multicolumn{2}{c}{\hspace{-0.4ex}\emph{sat}}&0.0&0.5&\ldots\\
		&      &     &&BaBb\tablefootmark{e}&    11.55&0.04           &    10.75&0.04    &    10.24&0.03    &     9.90&0.04      &      1.8&0.5            &      0.01  \\            
T\,27		& 0.78 & 125 &  13.3 & A &    11.27&0.03           &    10.48&0.04    &    10.08&0.02    &     9.38&0.03      &      0.9&0.3            &      0.50  \\            
		&      &     &       & B &    11.68&0.03           &    10.99&0.04    &    10.77&0.02    &    10.42&0.04      &      1.1&0.3            &      0.08  \\[0.5ex]     
T\,31		& 0.66 & 106 & 182.6 & A &     9.32&0.03           &     8.17&0.04    &     7.38&0.02    &     6.14&0.02      &      3.0&0.3            &      0.98  \\            
		&      &     &&BaBb\tablefootmark{e}&     9.74&0.03           &     8.83&0.04    &     8.51&0.02    &     8.24&0.04      &      1.8&0.5            &      0.04  \\[0.5ex]     
T\,33		& 2.40 & 384 & 284.4 & A &     9.65&0.03           &     8.01&0.03    &     6.68&0.02    &\multicolumn{2}{c}{\hspace{-0.4ex}\emph{sat}} & 4.3 & 0.3 &   \ldots  \\
		& 0.11 &  18 &  85.6 & Ba&    10.09&0.03           &     9.27&0.03    &     9.07&0.04    &\multicolumn{2}{c}{(\ldots)\tablefootmark{f}}&0.8&0.4            &     \ldots \\
		&      &     &       & Bb&    11.38&0.04           &    10.56&0.03    &    10.32&0.07    &\multicolumn{2}{c}{(\ldots)\tablefootmark{f}}&1.0&0.6            &     \ldots \\[0.5ex]     
T\,39		&1.25\tablefootmark{g}&200& 18.4& A &  10.37&0.03  &     9.67&0.03    &     9.48&0.03    &     9.88&0.03      &      0.0&0.4            &   $-$0.59  \\            
 		&4.50\tablefootmark{g}&720& 69.9& B &  10.77&0.03  &    10.00&0.03    &     9.83&0.03    &    10.31&0.03      &      0.4&0.4            &   $-$0.66  \\            
		&      &     &       & C &    11.01&0.04           &    10.29&0.03    &    10.09&0.04    &     9.75&0.03      &      0.2&0.5            &      0.14  \\[0.5ex]     
T\,41		& 0.78 & 125 & 324.3 & A &     7.61&0.03           &     7.35&0.04    &     7.28&0.02    &     7.36&0.01      &      0.0&0.3            &   $-$0.03  \\            
		&      &     &       & B &    11.25&0.28           &    10.57&0.13    &    10.27&0.06    &     9.93&0.04      &      0.3&2.5            &      0.09  \\[0.5ex]     
T\,43		& 0.78 & 126 & 352.0 & A &    11.56&0.03           &    10.32&0.03    &     9.55&0.02    &     8.64&0.03      &      5.5&0.3            &      0.45  \\            
		&      &     &       & B &    12.85&0.03           &    11.75&0.04    &    11.07&0.02    &    10.11&0.03      &      3.6&0.3            &      0.51  \\[0.5ex]     
T\,45		& 0.74 & 119 &  51.8 & A &    10.01&0.03           &     8.93&0.03    &     8.16&0.03    &     7.06&0.03      &      3.2&0.2            &      0.78  \\            
		&      &     &       & B &    11.78&0.03           &    10.99&0.03    &    10.59&0.03    &    10.10&0.04      &      2.3&0.2            &      0.11  \\[0.5ex]     
T\,51  		& 1.97 & 316 & 161.8 & A &     9.32&0.03           &     8.67&0.03    &     8.28&0.02    &     7.15&0.03      &      0.0&0.5            &      1.03  \\            
		&      &     &       & B &    11.39&0.03           &    10.72&0.03    &    10.38&0.03    &     9.66&0.03      &      0.0&0.5            &      0.53  \\[0.5ex]     
T\,54		& 0.24 &  38 & 248.0 & A &     8.65&0.04           &     8.24&0.03    &     8.24&0.06    &     8.19&0.01      &      0.0&0.5            &      0.02  \\            
		&      &     &       & B &    10.16&0.05           &     9.66&0.03    &     9.65&0.06    &     9.51&0.02      &      0.0&0.5            &      0.04  \\[0.5ex]     
\hline
\end{tabular}
\tablefoot{
  \tablefoottext{a}Separation uncertainties are $\Delta$sep\,=\,0\farcs01, $\Delta$PA\,=\,0.5$^\circ$. Projected separations in au assume a distance to Cha\,I of 160\,pc \citep{whi97}. For hierarchical triples, separations between a wide component and the close pair are given as well as the separation of the close components.
  \tablefoottext{b}Uncertainties are statistical errors from the differential photometry. Systematic uncertainties from relative aperture photometry are $\sim$0.1\,mag. 
  \tablefoottext{c}Due to a saturation of the reference star, no calibrated \Lp photometry could be measured. The magnitude difference is $\Delta L^\prime\!=\! L^\prime_B\! -\! L^\prime_A\!=\!4.69\pm0.19$
  \tablefoottext{d}These components are also known as Hn\,21\,E (primary) and Hn\,21\,W (secondary).
  \tablefoottext{e}These previously known visual binary components could not be resolved with our NACO adaptive optics observations. System photometry is listed. 
  \tablefoottext{f}These components are not separated in the \Lp-band images. The combined photometry returns $L^\prime_\mathrm{Ba+Bb}\!=\!8.60\pm0.03$.
  \tablefoottext{g}As no clear hierarchy is visible, sep.\ and PA of the secondary and tertiary components with respect to the primary are given.
}
}
\end{table*}

As suggested in the \emph{SINFONI data reduction cookbook}\footnote{\url{http://www.eso.org/sci/facilities/paranal/}\\\url{instruments/sinfoni/doc/}}, we removed a pattern of dark lines from all science and calibration files -- caused by a hard-coded bias-removal of the SINFONI instrument -- with the provided \emph{IDL} script. The subsequent reduction of all science and telluric reference frames used the standard \emph{SINFONI pipeline} in \emph{EsoRex}\footnote{\url{http://www.eso.org/sci/software/cpl/esorex.html}}. The pipeline executes bad-pixel removal, flat-fielding using a lamp flat, background subtraction, and wavelength calibration. A correction for differential atmospheric refraction was applied to exposures using the pre-optics 0.1 and 0.025. For each set of 4 or 16 dither exposures, the pipeline outputs a \emph{data cube} that consists of 2172 fully reduced and combined 2D images, each corresponding to a wavelength between 14\,750 and 24\,930\AA, in steps of 5\AA.

Before extracting the stellar 1D spectra from the data cubes, unusable pixel data (residual bad pixels, nearby ``negative stars'' from the sky subtraction, or blending with other stars in the image) were replaced by the median value computed in a two pixel-wide annulus around the defect at the wavelength they occur. The accuracy of this method was tested by applying the median substitution to good regions in a test cube. The difference between the extracted flux from the regular and the partly substituted cube was lower than 1\% at all wavelengths. 

All spectra were extracted twice with slightly different algorithms for different purposes: (a) to be compared with template spectra, which required a good recovery of the overall spectral shape, and (b) to measure \brgamma emission with a good signal-to-noise ratio (S/N).

\paragraph{(a) Extraction with preservation of the spectral slope. } Because adaptive optics correction produces a wavelength-dependent PSF, we applied a custom extraction routine. To accommodate the curved trace of each star throughout the cube (the position of the star between 1.5 and 2.4\,$\mu$m can differ by up to more than 1 spatial pixels) and the changing shape of the PSF as a function of wavelength, the center of each stellar component (target and standard) was traced and fitted by a fourth-order polynomial and its wavelength-dependent FWHM$_\lambda$ measured at every $\lambda$ by fitting a two-dimensional Moffat profile to the stellar disk. Multiple extractions were then conducted for each target component with aperture radii equal to 1, 2, 4, 6, 8, and 10 times the measured FWHM$_\lambda$. If the target PSF appeared to be strongly elliptical, or if two very close components of a multiple were extracted together, elliptical apertures were used, designed from an elliptical 2D Moffat profile fit. To make sure that the overall shape of the extracted spectrum was conserved, all extractions were compared the result from the largest aperture (10$\times$FWHM$_\lambda$), which contains close to 100\% of the flux. All extractions that agreed with this reference spectrum to better than 5\% at all wavelengths were then examined for their noise. The extraction with the highest S/N was selected (typically 2--4$\times$FWHM$_\lambda$).

\paragraph{(b) Extraction with maximum signal to noise. } To determine the \brgamma flux, only local correction is necessary. Thus, the overall spectral shape does not need to be conserved and the extraction can be optimized for S/N in the \brgamma region. This was achieved by using small extraction apertures (0.5--1$\times$FWHM$_\lambda$). The typical improvement in S/N over the continuum-conserving extraction at $\lambda_\mathrm{Br\upgamma}$ is a factor of $\sim$2. In the following, we refer to the results from (a) as \emph{flux-conserved extraction} and from (b) as \emph{high-S/N extraction}.

Standard-star spectra were extracted with the same procedures. To produce pure telluric spectra, convolved only with the instrumental profile, intrinsic absorption lines (except \brgamma) were removed by interpolating the continuum. Since a careful recovery of \brgamma flux is crucial for measuring accretion, this absorption line was removed according to the procedure described in \citet{dae12a}, which uses an estimate of the telluric spectrum around \brgamma from target spectra with neither absorption nor emission in this line. We corrected for the intrinsic continuum shape of the standard stars by dividing blackbody spectra with $T_{\rm eff}$ according to their spectral type. Telluric wavelenghts were fine-tuned by applying subpixel shifts after cross-correlation of 20-pixel-wide sections of the telluric and target spectra. 

For two of our target multiples, CHXR\,9C and CHXR\,71, two epochs of data with good S/N were taken (see Table~\ref{tab:obs:ChaIobservations}). Both data sets were reduced and extracted according to the scheme detailed above. Since no significant variability was apparent in the data, both extractions of all respective components were averaged to increase the S/N.

The final S/N of the reduced spectra depends on source brightness but never drops below 20 anywhere in the spectral windows of \H and \K-band for the shape-conserved reductions except for the two substellar companions. The typical S/N per pixel around \brgamma in the high-S/N reduction is around 80. The spectral region of low atmospheric transparency between 18000 and 20000\,\AA\ has a low S/N and was not used in the following. The final reduced and extracted spectra of all multiple star components are shown in Appendix~\ref{sec:app}, the most important stellar absorption features (see Table~\ref{tab:obs:features}) and \brgamma are marked. 

\begin{table}
  \caption[Spectral features identified in the observed spectra]{Spectral features identified in the observed spectra}
  \label{tab:obs:features}
  \centering
    \begin{tabular}{lccc}
      \hline\hline
      \multicolumn{1}{c}{$\lambda_\mathrm{c}$} & 
      \multicolumn{1}{c}{Width} &
      &
      \\
      \multicolumn{1}{c}{[\AA]} &
      \multicolumn{1}{c}{[\AA]} &
      \multicolumn{1}{c}{Species} &
      \multicolumn{1}{c}{Transition}\\[0.5ex]
      \hline
      14\,877.5     &         &   \ion{Mg}{I}       & $3d^3D^o_{1,2,3}$--$4f^3F^o_{2,3,4}$\\
      15\,025.0     &         &   \ion{Mg}{I}       & $4s^3S^o_1$--$4p^3P^o_2$            \\
      15\,040.2     &   *     &   \ion{Mg}{I}       & $4s^3S^o_1$--$4p^3P^o_1$            \\ 
      15\,168.4     &         &   \ion{K}{I}        & $3d^2D_{3/2}$--$4f^2F^o_{5/2}$      \\ 
      15\,740.7     &         &   \ion{Mg}{I}       & $4p^3P^o_0$--$4d^3D_1$              \\ 
      15\,765.8     &   *     &   \ion{Mg}{I}       & $4p^3P^o_2$--$4d^3D_3$              \\ 
      15\,888.4     &         &   \ion{Si}{I}       & $4s^1P^o_1$--$4p^1P_1$              \\ 
      16\,719.0     &         &   \ion{Al}{I}       & $^2P^o_{1/2}$--$^2D_{3/2}$          \\ 
      16\,750.5     &   *     &   \ion{Al}{I}       & $^2P^o_{3/2}$--$^2D_{5/2}$          \\ 
      17\,110.1     &         &   \ion{Mg}{I}       & $4s^1S_0$--$4p^1P^o_1$              \\[0.5ex] 
      \hline
    \end{tabular}
    \tablefoot{
      Characteristic photospheric features of late-type stars between 14500\,\AA\ and 18000\,\AA. Features between 20000 and 25000\,\AA\ are listed in Table~2 of \citet{dae12a}. Asterisks mark lines that blend with the line with the next shortest wavelength (the previous entry in the list). Transitions selected from \citet{cus05} and the \emph{NIST} Atomic Spectra Database (\url{http://www.nist.gov/pml/data/asd.cfm}).
}
\end{table}

\subsubsection{Flux calibration}
After extraction, all spectra were flux-calibrated with a routine similar to the one described in \citet{dae12a}. \Ks-band photometry was compared with synthetic photometry derived from convolving the uncalibrated SINFONI \HplusK spectra with a \Ks filter curve, which is superior to an extraction with \H or \HplusK filters in terms of noise and \brgamma calibration. Since the total flux of the high-S/N spectra is not conserved, these use the calibration factor derived from the flux-conserved extraction of the same target.

\section{Results\label{sec:ChaI:results}}
\subsection{Color-color diagram: photometric extinctions and color excesses\label{sec:ChaI:colorcolor}}
\begin{figure*}
  \centering
  \includegraphics[angle=0,width=0.9\textwidth]{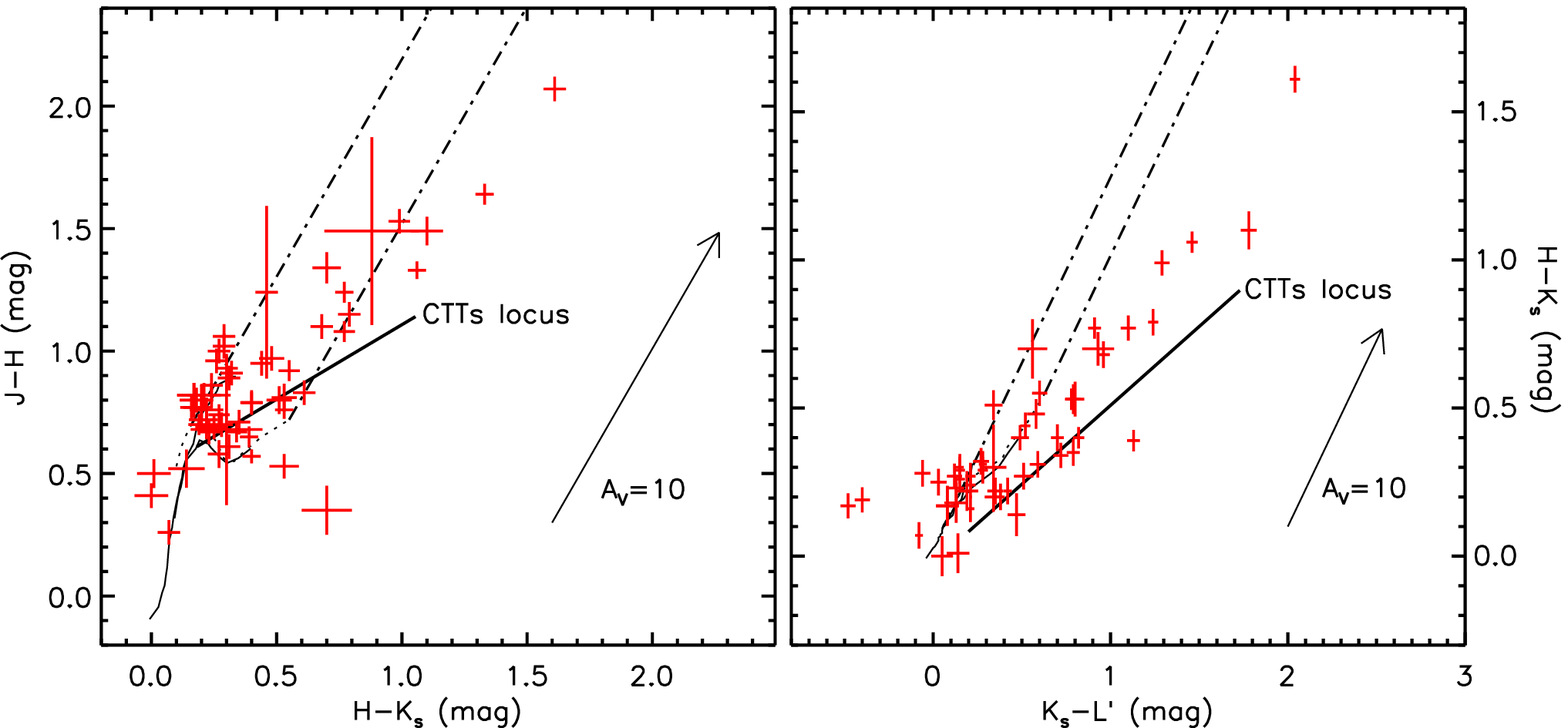}
  \caption[Color-color diagrams]{Color-color diagrams with photometric data from Table~\ref{tab:obs:ChaIphotometry}. Data are shown as red crosses, the giant and dwarf sequences (thin continuous lines) from \citet{bes88}, the CTTS locus \citep{mey97}, and a reddening vector of length $A_V$\,=\,10\,mag \citep{coh81} are overplotted. The dotted locus just above the dwarf sequence is the sequence derived from PMS stars in \citet{luh10}. Targets between the two dash-dotted lines can be dereddened to the PMS locus.}
  \label{fig:ChaI:colorcolordiagram}
\end{figure*}

Fig.~\ref{fig:ChaI:colorcolordiagram} shows the (\JH)--(\HKs) and (\HKs)--(\KsLp) color-color diagrams for Cha\,I binaries with photometric data from Table~\ref{tab:obs:ChaIphotometry}. The data are compared with the sequence of dwarfs and giants, the CTTS locus, and to a sequence of K4--M9 pre-main sequence stars taken from \citet{luh10}. The data show a clear clustering around the tip of the PMS sequence at a spectral type $\sim$M1.

The color-color diagrams were used to derive photometric extinctions. The extinction $A_V^\mathrm{phot}$ of a target component was determined by moving it toward bluer colors along the reddening vector in the (\JH)--(\HKs) diagram until intersecting either the CTTS locus or the PMS sequence. To decide whether to deredden to the CTTS locus or the PMS sequence, the (\HKs)--(\KsLp) diagram was used since here the CTTS locus is clearly separate from the PMS, dwarf, and giant sequences. Targets that are clearly associated with the CTTS locus in (\HKs)--(\KsLp), i.e., fall along it or cannot otherwise be dereddened to the PMS sequence, were dereddened to the corresponding CTTS locus in (\JH)--(\HKs). All other target components were dereddened to the PMS sequence. Any targets with \JH$\lesssim$\,0.5 were assigned an extinction of $A_V$=0. Uncertainties were determined from dereddening according to the uncertainties of the color measurements. The photometric extinctions $A_V^\mathrm{phot}$ are listed in Table~\ref{tab:obs:ChaIphotometry}.

Component extinctions of some of the wider binaries are published in \citet{luh07}, and values typically agree to within $\Delta A_V$\,$\approx$\,$\pm$1\,mag \citep[after conversion $A_J\!=\!0.276A_V$;][]{coh81}. 

Color excesses were derived for all targets according to
\begin{equation}
  E_{K_\mathrm{s}-L^\prime} = (K_\mathrm{s}-L^\prime)_\mathrm{obs} - 0.04A_V - (K_\mathrm{s}-L^\prime)_0
\end{equation}
with reference (\KsLp)$_0$ dwarf colors, converted to the 2MASS system, from \citet{bes88} according to the spectral types in Table~\ref{tab:ChaI:spectroscopy} (see Sect.~\ref{sec:ChaI:SpTs}). It is possible that the derived excesses over- or underestimate the real excess, since the near-IR colors of low-mass PMS stars and dwarfs of the same spectral type differ by up to a few times 0.1\,mag (compare e.g.\ \citealt{bes88} with \citealt{luh10}). Since there are no well-calibrated \KsLp values for pre-main sequence stars available in the literature, we estimated the size of this effect by comparing PMS $(K\!-\![3.6])$ colors \citep{luh10} to dwarf \KsLp colors and found a difference between 0.05 and 0.17\,mag in the spectral type range between K4 and M6. Although the \emph{Spitzer/IRAC} [3.6]-band and \Lp are not identical, this gives an indication of the magnitude of this effect. We assigned systematic uncertainties of 0.2\,mag to the derived $E_{K_\mathrm{s}-L^\prime}$ values in Table~\ref{tab:obs:ChaIphotometry}.

\subsection{Spectral types, extinction and veiling\label{sec:ChaI:SpTs}}
\begin{figure}
  \centering
  \includegraphics[angle=0,width=1.0\columnwidth]{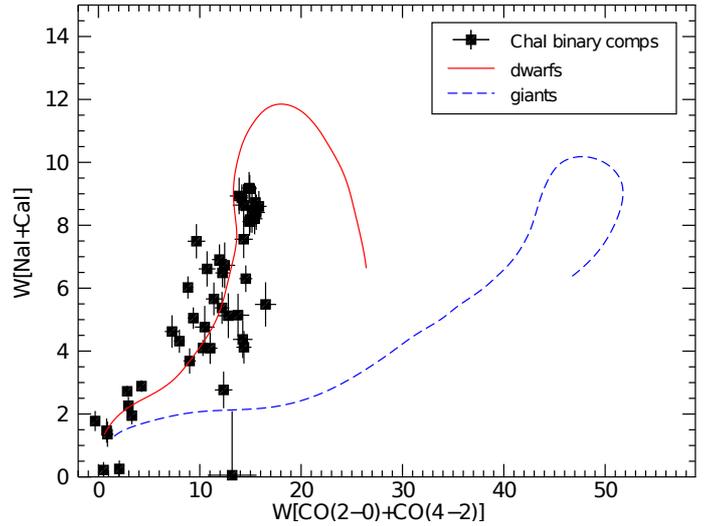}
  \caption[Comparison of Cha\,I target components to dwarf and giant spectra]{Comparison of the Cha\,I target components with dwarfs (red continuous) and giants (blue dashed) from the IRTF spectral library in terms of their equivalent widths in the mostly temperature sensitive \ion{Na}{I} and \ion{Ca}{I} lines and the surface-gravity-sensitive $^{12}$CO band heads. The dwarf and giant loci are derived from the IRTF spectral library \citep[see Fig.~4 in][]{dae12a}. Strong veiling reduces the measured equivalent widths and explains the clustering of targets close to the origin and to the left of the dwarf sequence.}
  \label{fig:ChaI:equivalentwidths}
\end{figure}

To measure spectral types, veiling, and to obtain another independent estimate of extinction, we compared the extracted flux-conserved spectra with dwarf spectra from the IRTF library in a similar way as described for ONC stars in \citet{dae12a}. Since Cha\,I is older than the ONC, the spectra in this sample will be even more similar to dwarf spectra in the IRTF library. Indeed, the strong concentration of Cha\,I targets along the dwarf sequence in a $W($\ion{Na}{I}$+$\ion{Ca}{I}$)$--$W($CO[2-0]$+$CO[4-2]$)$ diagram (Fig.~\ref{fig:ChaI:equivalentwidths}) indicates the similarity between the depth of the photospheric absorption features and thus the surface gravity $\log g$ between dwarfs and the Cha\,I PMS stars.
Accordingly, dwarf spectra were assumed to be suited to serve as templates and were used to derive spectral parameters.

\begin{figure}
  \centering
  \includegraphics[angle=0,width=1.0\columnwidth]{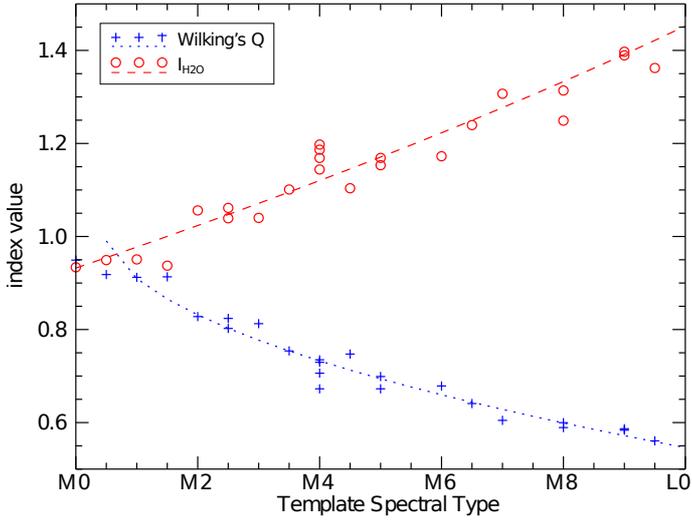}
  \caption[Wilking's Q and I$_\mathrm{H_2O}$ index]{Wilking's $Q$ and $I_\mathrm{H_2O}$ indices after \citet{wil99} and \citet{com00}, measured for the dwarf template sequence from the IRTF library \citep{cus05,ray09}. The dashed and dotted lines are best second-order polynomial fits to the data.}
  \label{fig:ChaI:WilkingsQ}
\end{figure}

To limit the number of templates that need to be tested for their compatibility with each target spectrum, we initially selected the spectral types following the extinction-independent \emph{Wilking's $Q$} \citep{wil99} and \emph{$I_{H_2O}$} indices \citep{com00}. Since our observations do not cover the whole wavelength range used by the original definition of $Q=F1/F2\,(F3/F2)^{1.22}$ [with $F1$ (2.07--2.13\,\mum), $F2$ (2.276--2.285\,\mum), and $F3$ (2.4--2.5\,\mum)], we redefined the last bin to be $F3^*$=2.4--2.43\,\mum. The introduced wavelength dependence of $Q$ is not significant in the limited spectral type range of the target sample, as was tested on template spectra of the same spectral types. A second-order polynomial was fit to both indices (Fig.~\ref{fig:ChaI:WilkingsQ}) derived from the templates as a function of spectral type assigned by the authors of the sequence \citep{cus05,ray09}.
The two derived relations for subclasses of M are
\begin{eqnarray}\label{eq:ChaI:waterspt}
  (\mathrm{SpT_M})_Q &=& 44.3 - 85.7\,Q + 41.9\,Q^2\quad,\\
  (\mathrm{SpT_M})_{I_{H_2O}} &=& -26.3 + 33.9\,I_{H_2O} - 6.18\,I_{H_2O}^2\quad,
\end{eqnarray}
with a residual scatter of $<$\,2 subclasses. Both indices become independent of spectral type for spectral types earlier than $\sim$M1 and were used only for determinations between M1 and L0.

$Q$ and $I_{H_2O}$ were calculated for the flux-conserved spectra of all targets. The two spectral type estimates, $(\mathrm{SpT_M})_Q$ and $(\mathrm{SpT_M})_{I_{H_2O}}$ which typically agree to within three subclasses, were averaged for each target. The subsequent detailed inspection of these and another $\pm$three subclasses was usually sufficient to find the best match between the spectral templates and the target component spectra. Targets with spectral types earlier than M1 were determined without this preselection.

The final best-fitting template was selected by analyzing the absolute and relative strength of multiple $T_\mathrm{eff}$-dependent absorption features in \H and \K-band (e.g.\, \ion{Na}{I}, \ion{Ca}{I}, \ion{Al}{I}, \ion{Mg}{I}, \ion{Si}{I}, see Table~\ref{tab:obs:features}) by visual comparison of the continuum-subtracted target spectra with the selection of IRTF spectra. The resulting spectral type uncertainties are typically between 0.5 and 2 subclasses, estimated from the range of possible spectral type matches. Attempts to use an automatic spectral type determination from line ratios did not return similarly good results. This is due to the weak strength of most absorption lines at the low spectral resolution of $R$\,$\approx$\,1400 and the comparably uncertain continuum determination in the near-IR spectra. 

The template with the best spectral type was modified by adding artificial reddening and veiling in three iterations, each with refined grids in extinction $A_V$, continuum excess $k$, and constant $c$ according to
\begin{equation}\label{eq:ONC:F}
  F^*(\lambda) = \left(\frac{F_\mathrm{phsph}(\lambda)}{c} + k\right)e^{-\tau_\lambda}
\end{equation}
\citep[cf.][]{pra03}. The best-fitting parameter combination and all uncertainties were derived through $\chi^2$ minimization \citep[for details see][]{dae12a}.

Owing to the relatively wide wavelength range of ($\sim$1.45--2.43\,\mum) of these observations, we tested whether the veiling can safely be assumed to be independent of $\lambda$. As suggested by observations of e.g.\ \citet{muz03}, veiling can take the shape of a blackbody at the dust sublimation temperature. To simulate this, the constant $k$ was multiplied with a blackbody curve of temperature 1500\,K and the fitting was repeated. The new best fits were at best indistinguishable from the fitting with constant excess, but mostly visibly worse, and the best $\chi^2$ was also typically larger. Consequently, $k$ was assumed to be constant for the derivation of $r_K$. The resulting best-fitting spectral types $A_V^\mathrm{spec}$ and $r_K$ are listed in Table~\ref{tab:ChaI:spectroscopy}.

\begin{figure}
  \centering
  \includegraphics[angle=0,width=1.0\columnwidth]{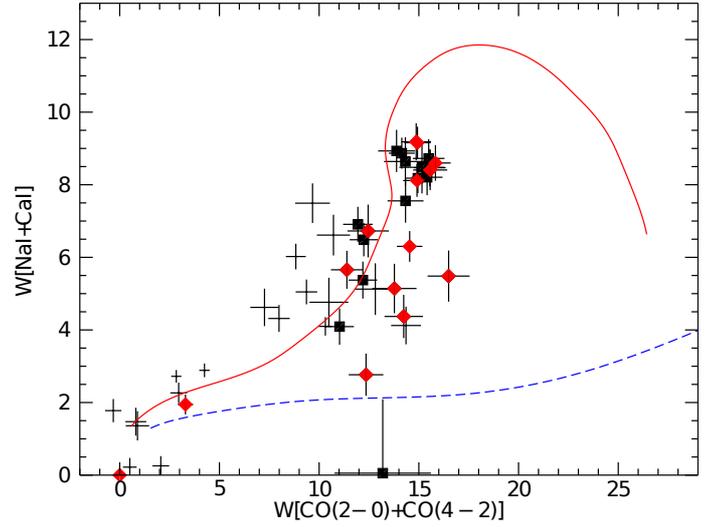}
  \caption[Comparison of Cha\,I target components to dwarf and giant spectra]{Same as Fig.~\ref{fig:ChaI:equivalentwidths}, but target components with a continuum significantly different from that of a dwarf star between 1.45 and 1.7\,\mum are marked with red diamonds. Low-veiling components with $r_K\!<\!0.1$ are marked with black squares and high-veiling components $r_K\!>\!0.1$ with error bars only.}
  \label{fig:ChaI:equivalentwidths2}
\end{figure}

Best-fit templates typically fit the target spectrum very well over the wavelength range between $\sim$1.7 and $\sim$2.43\,\mum. The wavelength range between $\sim$1.5 and $\sim$\,1.7\,\mum is only fit well in $\sim$70\% of the cases. Part of these deviations can be explained by uncertainties in the spectral extraction routine, which are larger at short wavelengths due to a lower Strehl ratio. The majority of deviations, however, are most likely due to broad water absorption features at 1.38\,\mum and 1.87\,\mum, which produce a characteristic triangular continuum shape in young late-type spectra \citep[e.g.,][]{sat92} that is not as pronounced in dwarf spectra because of their higher surface gravity.
That surface gravity is indeed the reason for the deviant continuum at $<$\,1.7\,\mum is demonstrated in Fig.~\ref{fig:ChaI:equivalentwidths2}, which shows that target components with the strongest bias toward the giant locus (i.e., with the lowest $\log g$, intermediate between that of dwarfs and giants) are almost exclusively stars that strongly deviate from dwarf spectra at $<$1.7\,\mum. 
To estimate the impact of a poor fit at wavelengths $<$\,1.7\,\mum on the parameter extraction, test fits were run in a limited wavelength range of 1.7--2.43\,\mum and compared with the fitting over the whole range. The changes in the measured $A_V$ and $r_K$ values were insignificant ($|\Delta A_V|\!<\!0.3$, $|\Delta r_K|\!<\!0.01$) with larger uncertainties, however, than when derived from the whole available wavelength range. Accordingly, dwarf templates were fit to all target components in the whole 1.45--2.43\,\mum range.

The early spectral type (B9) of T\,41\,A is outside the range covered by the spectral library in use. This value was copied from the \emph{simbad}\footnote{\url{http://simbad.u-strasbg.fr/simbad/}} database. No additional spectral values were derived for this component since discussing it is beyond the scope of this study. No spectral type could be derived for the brown dwarf candidates T\,14\,B and CHXR\,73\,B since the S/N of both spectra was too low to use spectral features for the determination of the spectral type. Assuming a spectral type of M9.5 for CHXR\,73\,B \citep{luh06}, however, spectral fitting could be ran, resulting in an extinction value of $A_V^\mathrm{spec}=1.6\pm0.7$ and veiling $r_K=0.02\pm0.17$.
The spectrum of T\,33\,A shows signs of strong extinction and veiling. From the shape of the continuum, it could be inferred that its spectral type is earlier than or equal to M3. Due to the strong veiling, however, photospheric features were not strong enough to further refine the SpT measurement. \citet{luh07} list T\,33\,A with a spectral type of G7, which the present data can neither confirm nor reject. The spectra of the components of ISO\,126 and T\,31 suffer from high veiling or extinction, or from both. Spectral fitting for these stars was impossible due to their weak absorption features. 

We compared our results with spectral types and extinctions derived by \citet{bra97}, who observed the individual components of T\,27, T\,31, and T\,39 with optical photometry and spectroscopy. Their component spectral types are mostly identical to our NIR-derived values or agree to within 1 (T\,31\,B) or 2 (T\,39\,A) subclasses. Their extinction values, however, are lower by on average 0.6\,mag for all target components except T\,39\,A. While this discrepancy can be explained by the subsequently assumed spectral type for this component, our systematically higher values might be a result of the insufficiently known dust properties and thus extinction laws in regions of star formation and circumstellar disks, which makes an accurate extrapolation of visual extinctions from near-IR measurements difficult \citep[see, e.g.,][]{nis09}. The relative extinction measurements (Sect.~\ref{sec:ChaI:relativeages}) are not affected by these systematics.

\begin{sidewaystable*}
\vspace{0.75\textheight}
{\tiny
{\scalefont{0.8}
\caption{Properties of individual binary and tertiary components}
\label{tab:ChaI:spectroscopy}
\centering
\begin{tabular}{lcr@{$\,\pm\,$}lr@{$\,\pm\,$}lr@{$\,\pm\,$}lr@{$\,\pm\,$}lr@{$\,\pm\,$}lr@{$\,\pm\,$}lr@{$\,\pm\,$}lr@{$\,\pm\,$}lr@{$\,\pm\,$}lcr@{$\,\pm\,$}lr@{$\,\pm\,$}l}
    \hline\hline
      & 
      &
      \multicolumn{2}{c}{} &
      \multicolumn{2}{c}{$T_\mathrm{eff}$} &
      \multicolumn{2}{c}{$A_V^\mathrm{spec}$} &
      \multicolumn{2}{c}{$r_K$} &
      \multicolumn{2}{c}{$L_*$} &
      \multicolumn{2}{c}{age} &
      \multicolumn{2}{c}{$M_*$} &
      \multicolumn{2}{c}{$R_*$} &
      \multicolumn{2}{c}{($W_{\mathrm{Br}\gamma}^\mathrm{accr}$)\tablefootmark{a}} &
      disk &
      \multicolumn{2}{c}{} &
      \multicolumn{2}{c}{$\dot{M}_\mathrm{acc}$} \\
      \multicolumn{1}{c}{Name} &
      Comp. &
      \multicolumn{2}{c}{SpT} &
      \multicolumn{2}{c}{$[K]$} &
      \multicolumn{2}{c}{[mag]} &
      \multicolumn{2}{r}{$(F_{K_\mathrm{ex}}/F_{K_{*}})$} &
      \multicolumn{2}{c}{$[L_\odot]$} &
      \multicolumn{2}{c}{$[Myr]$} &
      \multicolumn{2}{c}{$[M_\odot]$} &
      \multicolumn{2}{c}{$[R_\odot]$} &
      \multicolumn{2}{c}{[\AA]}&
      prob. &
      \multicolumn{2}{c}{log($L_\mathrm{acc}/L_\odot$)} &
      \multicolumn{2}{c}{$[10^{-9}M_\odot\mathrm{yr}^{-1}]$} \\
    \hline
B\,53      & A    & M2   & 1                 & 3560&150                &      0.1 & 0.4           & 0.00 & 0.09                      &  0.19  & 0.04           &      4.0 &$_{1.3}^{4.0}$  &  0.37 &$_{0.05}^{0.10}$              &               1.15 & 0.05 & \multicolumn{2}{c}{$<$0.91 } & 0.00 & \multicolumn{2}{c}{$<$$-$2.44} & \multicolumn{2}{c}{ $<$0.45}  \\       
           & B    & M4   &$_{1.5}^{0.5}$     & 3270&150                &      0.2 & 0.4           & 0.00 & 0.07                      &  0.05  & 0.01           &      8.0 &$_{3.2}^{7.0}$  &  0.20 &$_{0.05}^{0.07}$              &               0.71 & 0.03 & \multicolumn{2}{c}{$<$1.44 } & 0.00 & \multicolumn{2}{c}{$<$$-$2.96} & \multicolumn{2}{c}{ $<$0.16}  \\[0.5ex]
CHXR\,9C   & A    & M1.5 & 1                 & 3630&150                &      0.0 & 0.5           & 0.04 & 0.16                      &  0.20  & 0.03           &      4.4 &$_{1.6}^{5.1}$  &  0.41 &$_{0.06}^{0.10}$              &               1.14 & 0.05 & \multicolumn{2}{c}{$<$1.03 } & 0.00 & \multicolumn{2}{c}{$<$$-$2.34} & \multicolumn{2}{c}{ $<$0.51}  \\       
           & BaBb & M2   &$_{1.5}^{1}$       & 3560&180                &      0.1 & 0.5           & 0.05 & 0.16                      &  0.22  & 0.03           &      3.3 &$_{1.0}^{3.5}$  &  0.37 &$_{0.07}^{0.13}$              &               1.24 & 0.06 & \multicolumn{2}{c}{$<$0.87 } & 0.00 & \multicolumn{2}{c}{$<$$-$2.41} & \multicolumn{2}{c}{ $<$0.57}  \\[0.5ex]
CHXR\,15   & A    & M5   & 1                 & 3130&140                &      0.2 & 0.7           & 0.00 & 0.15                      &  0.06  & 0.01           &      4.5 &$_{0.9}^{2.5}$  &  0.16 &0.05                          &               0.87 & 0.04 & \multicolumn{2}{c}{$<$1.01 } & 0.01 & \multicolumn{2}{c}{$<$$-$2.88} & \multicolumn{2}{c}{ $<$0.29}  \\       
           & B    & M5   &$_{1}^{1.5}$       & 3130&170                &      0.1 & 0.6           & 0.00 & 0.13                      &  0.04  & 0.01           &      6.0 &$_{2.0}^{6.0}$  &  0.14 &0.06                          &               0.71 & 0.04 & \multicolumn{2}{c}{$<$1.13 } & 0.03 & \multicolumn{2}{c}{$<$$-$3.10} & \multicolumn{2}{c}{ $<$0.16}  \\[0.5ex]
CHXR\,28   & AaAb & K7   &$_{1}^{1.5}$       & 4060&290                &      2.7 & 0.4           & 0.02 & 0.13                      &  1.2   & 0.3            &      1.3 &$_{0.5}^{2.5}$  &  0.75 &0.25                          &               2.21 & 0.16 & \multicolumn{2}{c}{$<$1.01 } & 0.00 & \multicolumn{2}{c}{$<$$-$1.51} & \multicolumn{2}{c}{ $<$3.64}  \\       
           & B    & M2.5 &$_{0.5}^{1}$       & 3490&110                &      0.4 & 0.6           & 0.13 &$_{0.13}^{0.18}$           &  0.42  & 0.09           &      1.6 &0.4             &  0.36 &0.04                          &               1.78 & 0.06 & \multicolumn{2}{c}{$<$1.03 } & 0.01 & \multicolumn{2}{c}{$<$$-$2.01} & \multicolumn{2}{c}{ $<$1.93}  \\[0.5ex]
CHXR\,47   & A    & K5.5 &$_{1.5}^{1}$       & 4210&390                &      5.8 & 0.7           & 0.19 &$_{0.15}^{0.21}$           &  0.72  & 0.14           &      4.5 &$_{3.1}^{9.5}$  &  0.95 &$_{0.42}^{0.20}$              &               1.60 & 0.15 & \multicolumn{2}{c}{$<$1.76 } & 0.17 & \multicolumn{2}{c}{$<$$-$1.56} & \multicolumn{2}{c}{ $<$1.86}  \\       
           & B    & K7   &$_{0.5}^{1.5}$     & 4060&210                &      5.2 & 0.8           & 0.26 &$_{0.18}^{0.24}$           &  0.47  & 0.09           &      5.5 &$_{3.0}^{6.5}$  &  0.80 &$_{0.22}^{0.20}$              &               1.39 & 0.07 & \multicolumn{2}{c}{$<$1.78 } & 0.02 & \multicolumn{2}{c}{$<$$-$1.81} & \multicolumn{2}{c}{ $<$1.08}  \\[0.5ex]
CHXR\,49NE & A    & M2.5 &$_{0.5}^{1}$       & 3490&110                &      0.0 & 0.4           & 0.00 & 0.03                      &  0.20  & 0.04           &      3.4 &$_{0.8}^{1.2}$  &  0.34 &0.06                          &               1.22 & 0.04 & \multicolumn{2}{c}{$<$1.07 } & 0.01 & \multicolumn{2}{c}{$<$$-$2.46} & \multicolumn{2}{c}{ $<$0.50}  \\       
           & B    & M3.5 &$_{1}^{0.5}$       & 3340&110                &      0.0 & 0.3           & 0.00 & 0.04                      &  0.15  & 0.03           &      3.4 &$_{0.6}^{1.2}$  &  0.27 &0.06                          &               1.15 & 0.04 & \multicolumn{2}{c}{$<$1.12 } & 0.00 & \multicolumn{2}{c}{$<$$-$2.45} & \multicolumn{2}{c}{ $<$0.60}  \\[0.5ex]
CHXR\,68   & AaAb & M0   &1                  & 3850&180                &      0.3 & 0.5           & 0.00 & 0.11                      &  0.44  & 0.10           &      3.0 &$_{1.4}^{4.0}$  &  0.58 &$_{0.13}^{0.20}$              &               1.50 & 0.07 & \multicolumn{2}{c}{$<$1.40 } & 0.04 & \multicolumn{2}{c}{$<$$-$1.74} & \multicolumn{2}{c}{ $<$1.88}  \\       
           & B    & M1   &$_{0.5}^{1}$       & 3710&110                &      0.0 & 0.2           & 0.00 & 0.05                      &  0.12  & 0.02           &      12  &$_{7}^{17}$     &  0.46 &$_{0.18}^{0.10}$              &               0.83 & 0.02 & \multicolumn{2}{c}{$<$0.82 } & 0.00 & \multicolumn{2}{c}{$<$$-$2.79} & \multicolumn{2}{c}{ $<$0.12}  \\[0.5ex]
CHXR\,71   & A    & M3   &$_{1.5}^{0.5}$     & 3560&270                &      2.3 & 0.6           & 0.01 & 0.15                      &  0.11  & 0.03           &      7.0 &$_{3.5}^{8.0}$  &  0.37 &0.05                          &               0.90 & 0.03 & \multicolumn{2}{c}{$<$0.88 } & 0.00 & \multicolumn{2}{c}{$<$$-$2.77} & \multicolumn{2}{c}{ $<$0.16}  \\       
           & B    & M5   &$_{2}^{1}$         & 3130&270                &      1.5 & 0.6           & 0.04 & 0.18                      &  0.018 & 0.003          &      13  &$_{7}^{50}$     &  0.13 &$_{0.09}^{0.08}$              &               0.44 & 0.04 & \multicolumn{2}{c}{$<$1.05 } & 0.03 & \multicolumn{2}{c}{$<$$-$3.51} & \multicolumn{2}{c}{ $<$0.04}  \\[0.5ex]
CHXR\,73   & A    & M2.5 &$_{1}^{1.5}$       & 3490&180                &      1.6 & 0.7           & 0.02 & 0.17                      &  0.09  & 0.07           &      8   &$_{5}^{70}$     &  0.32 &0.11                          &               0.83 & 0.04 & \multicolumn{2}{c}{$<$0.98 } & 0.00 & \multicolumn{2}{c}{$<$$-$2.91} & \multicolumn{2}{c}{ $<$0.13}  \\       
           & B    & \multicolumn{2}{c}{\dots}&\multicolumn{2}{c}{\dots}&\multicolumn{2}{c}{\dots} & \multicolumn{2}{c}{\dots}        &\multicolumn{2}{c}{\dots}&\multicolumn{2}{c}{\dots}  & \multicolumn{2}{c}{\dots}            & \multicolumn{2}{c}{\dots} & \multicolumn{2}{c}{($<$4.48)}& 0.03 & \multicolumn{2}{c}{\dots}      & \multicolumn{2}{c}{\dots}     \\[0.5ex]
CHXR\,79   & A    & M2   &1                  & 3560&150                &      9.6 & 0.4           & 0.22 &$_{0.18}^{0.25}$           &  0.36  & 0.07           &      2.0 &$_{0.5}^{1.0}$  &  0.38 &$_{0.05}^{0.10}$              &               1.57 & 0.07 &           0.92 & 0.21        & 0.74 &            $-$1.53 & 0.24      &  4.88 &$_{3.66}^{3.83}$       \\       
           & B    & M4   &$_{1}^{0.5}$       & 3270&110                &      6.4 & 1.0           & 0.22 &$_{0.16}^{0.20}$           &  0.035 & 0.007          &      12  &$_{5.5}^{8}$    &  0.19 &$_{0.04}^{0.14}$              &               0.59 & 0.02 & \multicolumn{2}{c}{$<$1.96}  & 0.21 & \multicolumn{2}{c}{$<$$-$2.98} & \multicolumn{2}{c}{ $<$0.13}  \\[0.5ex]
Hn\,4      & A    & M4   &$_{1}^{0.5}$       & 3270&110                &      3.6 & 0.5           & 0.01 & 0.12                      &  0.10  & 0.02           &      4.2 &1.0             &  0.22 &$_{0.05}^{0.07}$              &               1.00 & 0.03 & \multicolumn{2}{c}{$<$1.61 } & 0.00 & \multicolumn{2}{c}{$<$$-$2.42} & \multicolumn{2}{c}{ $<$0.69}  \\       
           & B    & M3   &1                  & 3420&150                &      3.5 & 0.5           & 0.00 & 0.10                      &  0.11  & 0.02           &      5.2 &$_{0.7}^{5.0}$  &  0.30 &0.08                          &               0.95 & 0.04 & \multicolumn{2}{c}{$<$1.72 } & 0.00 & \multicolumn{2}{c}{$<$$-$2.39} & \multicolumn{2}{c}{ $<$0.52}  \\[0.5ex]
Hn\,21     & A    & M3.5 &$_{1}^{0.5}$       & 3340&110                &      2.6 & 0.7           & 0.02 & 0.14                      &  0.11  & 0.02           &      4.5 &$_{1.0}^{2.0}$  &  0.26 &0.05                          &               0.99 & 0.03 & \multicolumn{2}{c}{$<$0.99 } & 0.18 & \multicolumn{2}{c}{$<$$-$2.81} & \multicolumn{2}{c}{ $<$0.24}  \\       
           & B    & M6.5 &$_{2.5}^{2}$       & 2940&200                &      0.0 & 0.6           & 0.00 & 0.14                      &  0.030 & 0.006          &      5.5 &$_{1.5}^{4.5}$  &  0.08 &$_{0.05}^{0.10}$ \tablefootmark{c}&           0.66 & 0.04 & \multicolumn{2}{c}{$<$1.05 } & 0.01 & \multicolumn{2}{c}{$<$$-$3.22} & \multicolumn{2}{c}{ $<$0.20}  \\[0.5ex]
ISO\,126   & A    & M0.5 &$_{2}^{1.5}$       & 3780&400                &\multicolumn{2}{c}{$>$4}  &\multicolumn{2}{c}{$\gg$1}        &  0.50  & 0.09           &      2.0 &$_{1.1}^{8}$    &  0.52 &$_{0.19}^{0.38}$              &               0.47 & 0.05 & \multicolumn{2}{c}{($<$0.75)}& 0.00 & \multicolumn{2}{c}{\dots     } & \multicolumn{2}{c}{\dots}     \\       
           & B    & M0.5 &$_{2}^{1.5}$       & 3780&400                &\multicolumn{2}{c}{$>$4}  &\multicolumn{2}{c}{$\gg$1}        &  0.090 & 0.006          &   (23 &$_{18}^{\dots}$)   &  0.50 &0.2                           &               0.70 & 0.07 & \multicolumn{2}{c}{($<$0.65)}& 0.00 & \multicolumn{2}{c}{\dots     } & \multicolumn{2}{c}{\dots}     \\[0.5ex]
T\,3       & A    & M1.5 &$_{2.5}^{1.5}$     & 3630&320                &      2.2 & 1.1           & 1.3  &$_{0.6 }^{1.8 }$           &  0.50  & 0.09           &      1.6 &$_{0.8}^{2.3}$  &  0.42 &$_{0.12}^{0.21}$              &               1.79 & 0.16 &           6.52 & 0.28        & 1.00 &            $-$0.62 & 0.05      & 40.9  &$_{13.2}^{21.4}$       \\       
           & B    & M3   &$_{2}^{1.5}$       & 3420&250                &      0.5 & 0.7           & 0.18 &$_{0.14}^{0.22}$           &  0.10  & 0.02           &      5.8 &$_{2.5}^{10}$   &  0.29 &$_{0.10}^{0.13}$              &               0.91 & 0.07 & \multicolumn{2}{c}{$<$1.37}  & 0.00 & \multicolumn{2}{c}{$<$$-$2.45} & \multicolumn{2}{c}{ $<$0.45}  \\[0.5ex]
T\,6       & A    & K1   &$_{2}^{3}$         & 5080&490                &      0.0 & 0.9           & 1.5  &$_{0.7}^{2.1}$             &  1.6   & 0.4            &      10  &$_{7}^{12}$     &  1.4  &0.2                           &               1.65 & 0.16 & \multicolumn{2}{c}{$<$0.85}  & 0.00 & \multicolumn{2}{c}{$<$$-$1.37} & \multicolumn{2}{c}{ $<$2.01}  \\       
           & B    & M6   &$_{3.5}^{1}$       & 2990&320                &      4.6 & 1.2           & 0.54 & 0.14                      &  0.06  & 0.01           &      4.2 &$_{4.0}^{5.8}$  &  0.10 &$_{0.07}^{0.11}$              &               0.89 & 0.09 & \multicolumn{2}{c}{$<$2.00}  & 0.00 & \multicolumn{2}{c}{$<$$-$2.55} & \multicolumn{2}{c}{ $<$1.00}  \\[0.5ex]
T\,14      & A    & K7   &$_{0.5}^{1}$       & 4060&180                &      0.1 & 0.5           & 0.33 &$_{0.20}^{0.28}$           &  0.67  & 0.15           &      3.0 &$_{1.5}^{3.0}$  &  0.78 &0.20                          &               1.66 & 0.07 &           5.86 & 0.33        & 0.00 &            $-$0.90 & 0.04      & 10.7  &3.0                    \\       
           & B    & \multicolumn{2}{c}{\dots}&\multicolumn{2}{c}{\dots}&\multicolumn{2}{c}{\dots} & \multicolumn{2}{c}{\dots}        &\multicolumn{2}{c}{\dots}&\multicolumn{2}{c}{\dots}  & \multicolumn{2}{c}{\dots}            & \multicolumn{2}{c}{\dots} & \multicolumn{2}{c}{\dots   } & 0.00 & \multicolumn{2}{c}{\dots}      & \multicolumn{2}{c}{\dots}     \\[0.5ex]
T\,26      & A    & G0   &$_{3}^{1.5}$       & 6030&290                &      6.6 & 0.9           & 0.18 &$_{0.10}^{0.14}$           &  4.7   & 1.0            &      7.1 &$_{0.7}^{3.0}$  &  1.7  &0.2                           &               2.00 & 0.10 &           5.62 & 0.30        & 1.00 &            $-$0.50 & 0.15      & 14.9  &6.4                    \\       
           & BaBb & M5   &$_{2.5}^{1.0}$     & 3130&290                &      0.9 & 0.8           & 0.00 & 0.16                      &  0.11  & 0.02           &      3.0 &$_{2.8}^{3.0}$  &  0.18 &$_{0.08}^{0.06}$              &               1.13 & 0.11 & \multicolumn{2}{c}{$<$1.18 } & 0.01 & \multicolumn{2}{c}{$<$$-$2.49} & \multicolumn{2}{c}{ $<$0.81}  \\[0.5ex]
T\,27      & A    & M1.5 &$_{0.5}^{2}$       & 3630&180                &      2.9 & 0.6           & 0.10 & 0.19                      &  0.14  & 0.03           &      7.0 &$_{3.0}^{9.0}$  &  0.40 &$_{0.06}^{0.14}$              &               0.96 & 0.05 &           1.83 & 0.22        & 1.00 &            $-$2.07 & 0.25      &  0.82 &$_{0.70}^{0.65}$       \\       
           & B    & M3   &1                  & 3420&150                &      2.5 & 0.7           & 0.00 & 0.12                      &  0.09  & 0.02           &      6.4 &$_{2.2}^{4.7}$  &  0.29 &0.07                          &               0.87 & 0.04 & \multicolumn{2}{c}{$<$1.00 } & 0.01 & \multicolumn{2}{c}{$<$$-$2.85} & \multicolumn{2}{c}{$<$0.17}   \\[0.5ex]
T\,31      & A    & K7   &$_{1.5}^{3}$       & 4060&460                &\multicolumn{2}{c}{\dots} &\multicolumn{2}{c}{($\gg$1)}      &  1.6   & 0.3            &      1.0 &$_{0.7}^{3.0}$  &  0.75 &$_{0.35}^{0.50}$              &               2.55 & 0.29 &          (9.35 & 0.19)       & 1.00 & \multicolumn{2}{c}{$>$$-$1.51} & \multicolumn{2}{c}{$>$4.20}   \\       
           & BaBb & M0   &1.5                & 3850&290                &\multicolumn{2}{c}{\dots} &\multicolumn{2}{c}{($\sim$0)}     &  0.76  & 0.10           &      1.5 &$_{0.6}^{2.5}$  &  0.57 &$_{0.19}^{0.28}$              &               1.97 & 0.15 & \multicolumn{2}{c}{($<$0.87)}& 0.01 & \multicolumn{2}{c}{\dots     } & \multicolumn{2}{c}{\dots}     \\[0.5ex]
T\,33      & A    &\multicolumn{2}{c}{$<$M3} &\multicolumn{2}{c}{\dots}&\multicolumn{2}{c}{\dots} &\multicolumn{2}{c}{($\gg$1)}      &\multicolumn{2}{c}{\dots}&\multicolumn{2}{c}{\dots}  &  \multicolumn{2}{c}{\dots}           & \multicolumn{2}{c}{\dots} & \multicolumn{2}{c}{($<$0.84)}& 0.00 & \multicolumn{2}{c}{\dots     } & \multicolumn{2}{c}{\dots}     \\       
           & BaBb & M0.5 &$_{1.5}^{1}$       & 3780&210                &      1.5 & 0.5           & 0.00 & 0.12                      &  0.57  & 0.13           &      1.8 &$_{0.7}^{2.2}$  &  0.52 &$_{0.14}^{0.18}$              &               1.77 & 0.10 & \multicolumn{2}{c}{$<$0.99 } & 0.00 & \multicolumn{2}{c}{$<$$-$1.83} & \multicolumn{2}{c}{ $<$2.01}  \\[0.5ex]
T\,39      & A    & M2.5 &0.5                & 3490&70                 &      0.2 & 0.4           & 0.00 & 0.08                      &  0.25  & 0.05           &      2.7 &$_{0.5}^{0.7}$  &  0.36 &0.05                          &               1.37 & 0.03 & \multicolumn{2}{c}{$<$1.02 } & 0.00 & \multicolumn{2}{c}{$<$$-$2.17} & \multicolumn{2}{c}{ $<$1.03}  \\       
           & B    & M2   &0.5                & 3560&70                 &      0.2 & 0.4           & 0.00 & 0.07                      &  0.20  & 0.04           &      4.0 &$_{1.0}^{1.3}$  &  0.38 &0.05                          &               1.17 & 0.02 & \multicolumn{2}{c}{$<$1.05 } & 0.00 & \multicolumn{2}{c}{$<$$-$2.33} & \multicolumn{2}{c}{ $<$0.58}  \\       
           & C    & M2.5 &0.5                & 3490&70                 &      0.0 & 0.3           & 0.00 & 0.07                      &  0.14  & 0.03           &      4.7 &$_{1.2}^{2.3}$  &  0.34 &0.05                          &               1.04 & 0.02 & \multicolumn{2}{c}{$<$1.12 } & 0.01 & \multicolumn{2}{c}{$<$$-$2.42} & \multicolumn{2}{c}{ $<$0.47}  \\[0.5ex]
T\,41      & A  &\multicolumn{2}{c}{B9\tablefootmark{b}}&\multicolumn{2}{c}{10500}&\multicolumn{2}{c}{\dots} & \multicolumn{2}{c}{\dots}        &\multicolumn{2}{c}{\dots}&\multicolumn{2}{c}{\dots}  & \multicolumn{2}{c}{\dots} & \multicolumn{2}{c}{\dots} & \multicolumn{2}{c}{$<$0.36 } & 0.00 & \multicolumn{2}{c}{\dots}      & \multicolumn{2}{c}{\dots}     \\       
           & B    & M3.5 &0.5                & 3340&70                 &      0.0 & 0.2           & 0.00 & 0.07                      &  0.11  & 0.08           &      4.5 &$_{1.9}^{15}$   &  0.26 &0.07                          &               0.99 & 0.02 & \multicolumn{2}{c}{$<$1.28 } & 0.00 & \multicolumn{2}{c}{$<$$-$2.44} & \multicolumn{2}{c}{ $<$0.55}  \\[0.5ex]
T\,43      & A    & M3.5 &$_{2}^{1.0}$       & 3340&290                &      3.4 & 0.7           & 0.44 & 0.16                      &  0.29  & 0.06           &      1.8 &$_{1.6}^{1.2}$  &  0.30 &$_{0.10}^{0.11}$              &               1.62 & 0.13 & \multicolumn{2}{c}{$<$1.07 } & 0.00 & \multicolumn{2}{c}{$<$$-$2.34} & \multicolumn{2}{c}{ $<$0.99}  \\       
           & B    & M7   &$_{2.5}^{1.0}$     & 2880&360                &      4.5 & 0.8           & 0.00 & 0.16                      &  0.05  & 0.01           &      4.0 &$_{3.9}^{4.0}$  &  0.08 &$_{0.07}^{0.12}$ \tablefootmark{c} &          0.91 & 0.11 & \multicolumn{2}{c}{$<$1.07 } & 0.00 & \multicolumn{2}{c}{$<$$-$3.00} & \multicolumn{2}{c}{ $<$0.46}  \\[0.5ex]
T\,45      & A    & M1   &$_{1.5}^{1}$       & 3710&200                &      2.8 & 1.2           & 1.5  &$_{0.7}^{2.2}$             &  0.81  & 0.15           &      1.1 &$_{0.3}^{0.6}$  &  0.49 &0.12                          &               2.18 & 0.12 &          12.62 & 0.47        & 1.00 &            $-$0.26 & 0.10      & 98  &35                       \\       
           & B    & M5   &$_{2.5}^{1}$       & 3130&210                &      1.6 & 0.9           & 0.22 &$_{0.16}^{0.19}$           &  0.10  & 0.02           &      3.3 &$_{2.9}^{2.7}$  &  0.18 &$_{0.06}^{0.05}$              &               1.08 & 0.07 & \multicolumn{2}{c}{$<$1.28 } & 0.01 & \multicolumn{2}{c}{$<$$-$2.62} & \multicolumn{2}{c}{ $<$0.58}  \\[0.5ex]
T\,51      & A    & K7   &3                  & 4060&590                &      1.1 & 0.7           & 0.48 &$_{0.23}^{0.40}$           &  0.76  & 0.17           &      2.5 &$_{1.7}^{13}$   &  0.78 &0.40                          &               1.77 & 0.26 & \multicolumn{2}{c}{$<$1.08 } & 0.01 & \multicolumn{2}{c}{$<$$-$1.55} & \multicolumn{2}{c}{ $<$2.56}  \\       
           & B    & M2.5 &1.5                & 3490&220                &      3.8 & 0.8           & 0.10 & 0.20                      &  0.10  & 0.02           &      7.2 &$_{3.4}^{12}$   &  0.32 &$_{0.09}^{0.12}$              &               0.85 & 0.05 &           1.11 & 0.38        & 0.68 &            $-$2.67 & 0.47      &  0.23 &0.45                   \\[0.5ex]
T\,54      & A    & G8   &3                  & 5507&440                &      0.2 & 0.5           & 0.00 & 0.10                      &  2.0   & 0.4            &      16  &$_{10}^{30}$    &  1.3  &0.2                           &               1.52 & 0.12 & \multicolumn{2}{c}{$<$0.87 } & 0.48 & \multicolumn{2}{c}{$<$$-$1.69} & \multicolumn{2}{c}{ $<$0.96}  \\       
           & B    & K7   &1.5                & 4060&310                &      0.0 & 0.5           & 0.04 & 0.13                      &  0.35  & 0.08           &      9.0 &$_{6.5}^{16}$   &  0.80 &$_{0.3}^{0.15}$               &               1.20 & 0.09 & \multicolumn{2}{c}{$<$1.07 } & 0.00 & \multicolumn{2}{c}{$<$$-$2.23} & \multicolumn{2}{c}{ $<$0.35}  \\[0.5ex]
\hline
\end{tabular}   
\tablefoot{
  \tablefoottext{a}Spectral type from \emph{Simbad}.
  \tablefoottext{b}Values in parantheses mark lower limits since no veiling could be measured.
  \tablefoottext{c}Masses $<$0.1\,M$_\odot$\ lie outside the scope of the \citet{sie00} models. Values are estimated through extrapolation.
}
}
}
\end{sidewaystable*}

\subsubsection[Photometric vs.\ spectroscopic extinctions]{Photometric vs.\ spectroscopic extinctions}
\begin{figure}
  \centering
  \includegraphics[angle=0,width=0.79\columnwidth]{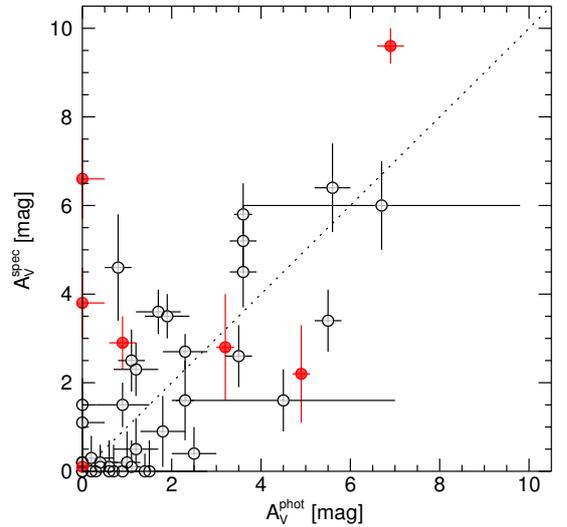}
  \caption[Spectroscopic versus photometric extinctions]{Spectroscopic versus photometric extinctions of all Cha\,I multiple star components for which both quantities could be measured. Red filled circles mark accreting components, all other components are marked with open circles. The dotted line shows equal extinctions.}
  \label{fig:ChaI:AVspec_vs_AVphot}
\end{figure}

The spectroscopic measurements provide an independent estimate of the interstellar extinction in addition to the photometric derivation in Sect.~\ref{sec:ChaI:colorcolor}. Fig.~\ref{fig:ChaI:AVspec_vs_AVphot} shows a comparison between both sets of extinction values.
While the scatter around the line of equal extinctions is relatively high -- deviations of 3\,mag and more do appear -- no systematic differences could be noticed. There is also no systematic difference between accreting components and non-accretors visible. The agreement between photometric and spectroscopic values indicates that the other spectroscopically inferred parameters (spectral type, veiling) are also on average compatible with the photometric results. Since the photometric determination of extinction, however, relies on shorter and thus more extinction-sensitive wavelengths (\JH color) than our spectroscopy (\HplusK), we used the photometrically derived values of all target components in the following.

\subsection{Effective temperatures and luminosities: the HR-diagram}
\begin{figure}
  \centering
  \includegraphics[angle=0,width=0.88\columnwidth]{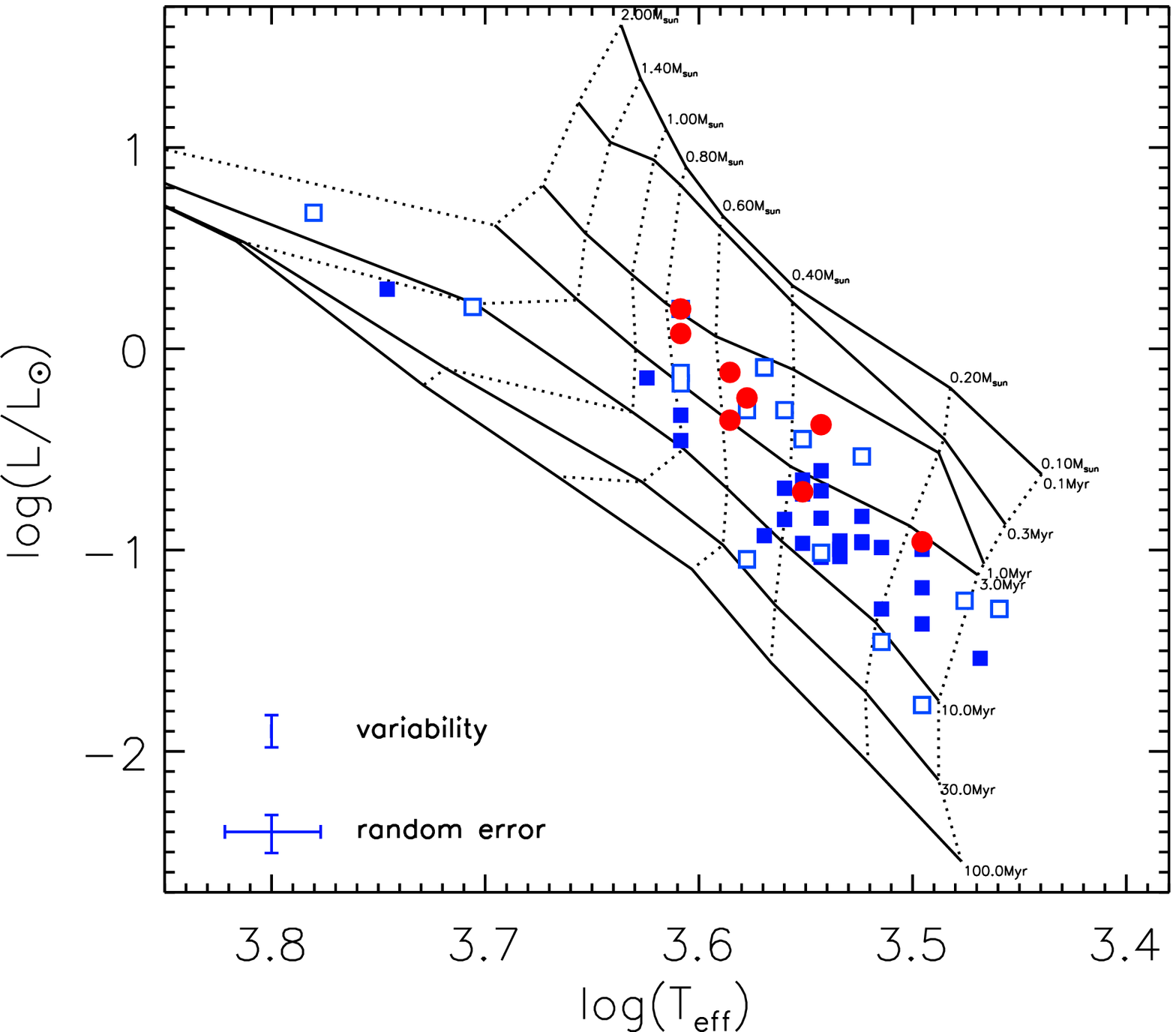}
  \caption[HR diagram of Cha~I target components with models of Siess et al.]{HR diagram with the isochrones of \citet{sie00}. Filled red circles show spatially unresolved components, either close binary components of visual triple stars or spectroscopic binaries. Luminosities of these targets probably appear to be higher than single stars at the same $T_\mathrm{eff}$. Open blue squares show targets with strong near-IR excess (E(\KsLp)\,$>$\,0.5 or \HKs$\,>$\,0.6) and filled blue squares all other components of the Chamaeleon\,I sample. In the lower left corner we present the average random error and the luminosity uncertainty introduced by 0.2\,mag variability in \J-band (estimated from measurements, see Sect.\ref{sec:obs:ChaIphotometry}).}
  \label{fig:ChaI:HRdiagram}
\end{figure}

Spectral types were converted to effective temperatures according to \citet{sch82} (earlier than M0) and \citet{luh03} (later than or equal to M0). The luminosity of each target component was estimated from its absolute $M_J$ magnitude \citep[assuming a distance to Cha\,I of 160$\pm$15\,pc;][]{whi97} and bolometric corrections $BC\!_J$ \citep{har94}. The luminosity uncertainties were calculated from the propagation of the uncertainties of all input parameters; they are dominated by the uncertainty in the distance measurement. Luminosities, effective temperatures, and stellar radii are listed in Table~\ref{tab:ChaI:spectroscopy} and are used to compile the Hertzsprung-Russell diagram shown in Fig.~\ref{fig:ChaI:HRdiagram}.

\subsubsection{Stellar masses and ages\label{sec:ChaI:masses}}
\begin{figure}
  \centering
  \includegraphics[angle=0,width=0.88\columnwidth]{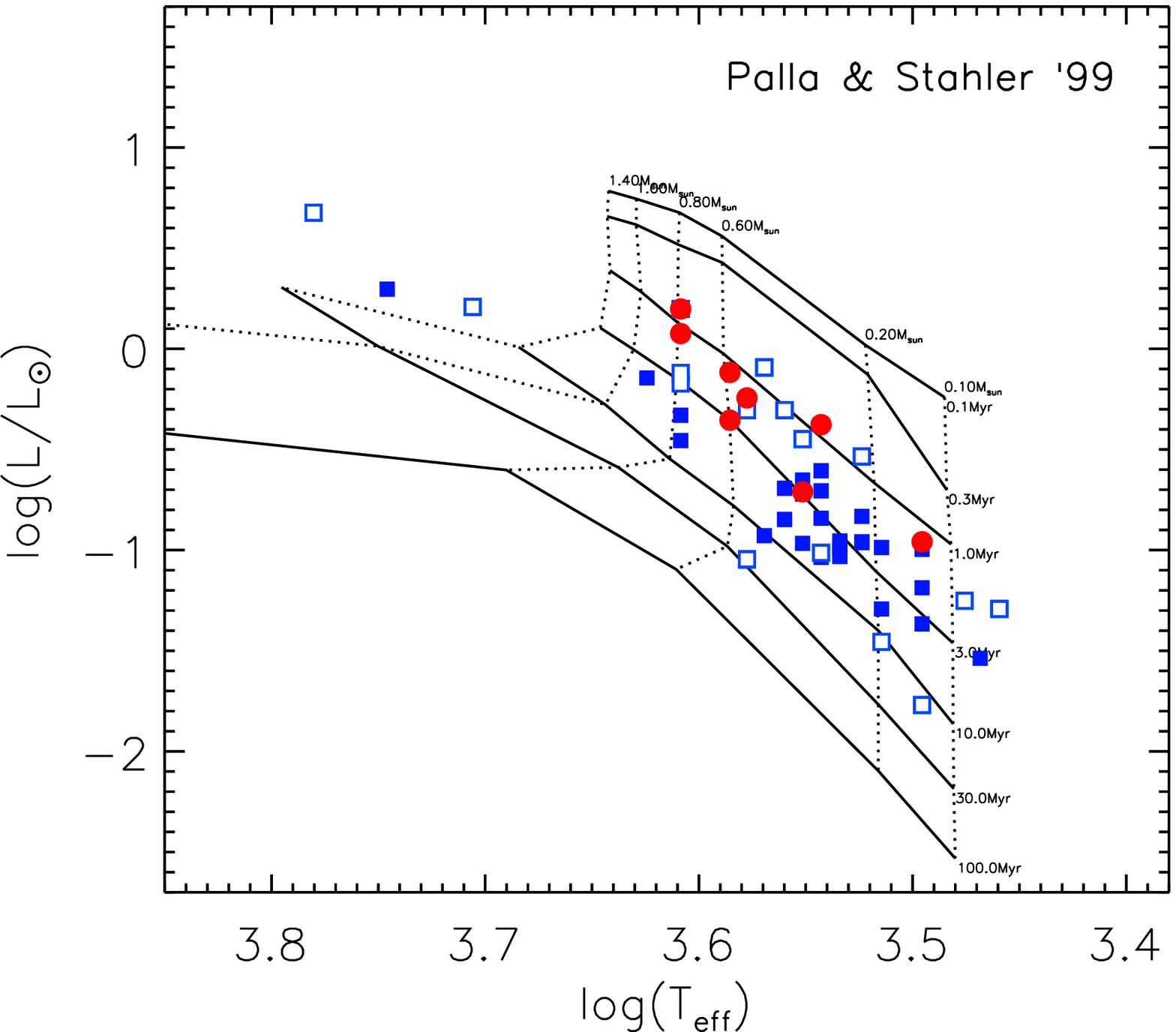}
  \caption[HR diagram of Cha~I target components with models of Palla \& Stahler]{Same as Fig.~\ref{fig:ChaI:HRdiagram}, but with isochrones of \citet{pal99}.}
  \label{fig:ChaI:HRdiagramPalla}
\end{figure}

The HR-diagram enables us to derive ages and masses of all components when comparing target locations with model isochrones. To derive the age and mass values in Table~\ref{tab:ChaI:spectroscopy}, models of \citet{sie00} were used. Because all derived values depend on the choice of model for the applied isochrones, a comparison of the same data with \citet{pal99} isochrones is shown in Fig.~\ref{fig:ChaI:HRdiagramPalla}.
Qualitatively, there seems to be a slight bias toward younger ages in these latter models, because the sequence of low-color excess target components follows the 3\,Myr isochrone of \citeauthor{pal99} closely, while the same targets lie mostly below the 3\,Myr isochrone of the \citet{sie00} models in Fig.~\ref{fig:ChaI:HRdiagram} (see Sect.~\ref{sec:ChaI:relativeages} for a discussion of the derived component ages). Moreover, masses below $\sim$0.2\,M$_\odot$\ appear on average to be slightly less massive when derived with the \citeauthor{pal99} models instead of those by \citet{sie00}. In the following, the derived ages and masses from the \citeauthor{sie00} isochrones were used to guarantee comparability with the ONC results from \citet{dae12a}, where the same model was applied. The differences between the inferred parameters from both sets of isochrones are probably smaller than their intrinsic systematic uncertainties. While most results were only used for order of magnitude calculations, the inferred \emph{mass ratios}, which is discussed in Sect.~\ref{sec:ChaI:differentialdiskevolution}, have been shown to be quite insensitive to the choice of isochrone \citep[e.g.,][]{cor13,laf08a}.

As expected, Fig.~\ref{fig:ChaI:HRdiagram} shows that unresolved binary stars have higher-than-average luminosities compared with individually observed components at the same $T_\mathrm{eff}$. Depending on the brightness ratio of these close components, the derived ages can appear artificially young. The same is true for stars with strong near-IR excess. While excess is typically assumed to be weak in the \J-band \citep[see, e.g.,][]{mey97}, stars with overall strong excess emission can exhibit enough excess flux in \J-band to significantly increase the estimated bolometric luminosity \citep{fis11}. Additionally, extinction might be overestimated, as discussed in Sect.~\ref{sec:ChaI:relativeages}, which can also lead to higher derived luminosities. These effects explain why the ten stars with the youngest derived ages are all either unresolved binaries or stars with strong color excess.

\subsection{Accretion\label{sec:ChaI:accretion}}
Accretion activity was inferred from the existence of \brgamma (21\,665\,\AA) emission. To assess the strength of this line, we used the high-S/N spectral extraction.

\subsubsection[Equivalent widths of Br-gamma emission]{Equivalent widths \Wbrgamma}
Equivalent widths were integrated in a 56\AA-wide interval around the centered \brgamma peak\footnote{Absorption features are defined to have negative equivalent width values and emission is positive.}. The continuum and noise background were estimated from regions shortward and longward of the line's central wavelength. \brgamma measurements were performed on the calibrated spectra, correcting for veiling by applying $W_\lambda = (1+r_K)\times W_\lambda^\mathrm{measured}$ \citep{dae12a}. To correct for the intrinsic \brgamma absorption of spectral types earlier than M1, we measured the equivalent width in the same wavelength range for the spectral sequence of dwarf stars in the IRTF database.
\begin{figure}
  \centering
  \includegraphics[angle=0,width=1.0\columnwidth]{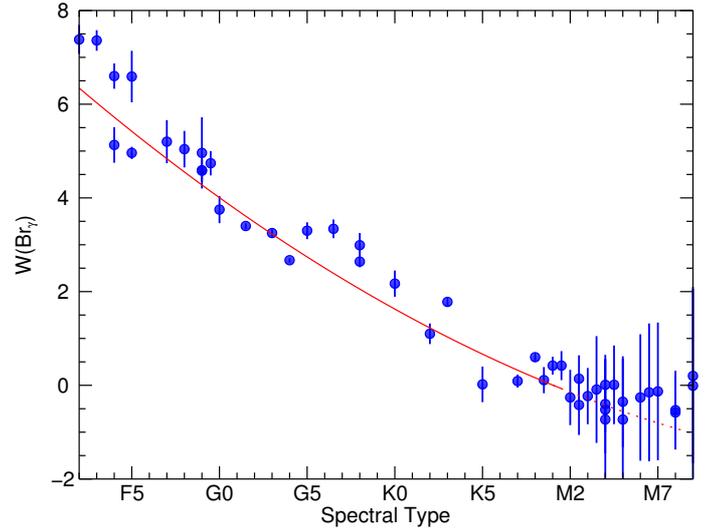}
  \caption[Br-gamma absorption strength in dwarfs]{Equivalent widths of \brgamma absorption lines as a function of spectral type measured from the IRTF dwarf template sequence. The continuous red line shows our best fit (Eq.~\ref{eq:ChaI:brgammaabsorption}).}
  \label{fig:ChaI:brgammaabsorption}
\end{figure}
The correlation between \Wbrgamma and spectral type is shown in Fig.~\ref{fig:ChaI:brgammaabsorption} and was fit by a quadratic polynomial
\begin{eqnarray}
  W_\mathrm{Br\upgamma}^\mathrm{photosph}(\mathrm{SpT}) = &&(0.16\pm0.05) - (0.16\pm0.01)\times\mathrm{SpT} \nonumber\\
                                                          &+& (0.0030\pm0.0004)\times\mathrm{SpT}^2\quad,\label{eq:ChaI:brgammaabsorption}
\end{eqnarray}
valid in a spectral type range between F0 and M1, with spectral types (SpT) measured as M0=0, M5=5, K0=$-$8, and so forth\footnote{The adopted classification omits spectral types K8 and K9}. The absorption for spectral types later than M1 is low and no correction is applied here. Although a rather large correction is probably necessary, spectral types earlier than F0 remained uncorrected because the spectral template sequence does not extend to the earliest spectral type of this sample (T41\,A, B9). The correction terms, tabulated in Table~\ref{tab:ChaI:brgammaabsorption}, were applied as
\begin{equation}
  W_\mathrm{Br\upgamma}^\mathrm{accr} = W_\lambda-W_\mathrm{Br\upgamma}^\mathrm{photosph}\quad,
\end{equation}
resulting in equivalent widths $W_\mathrm{Br\upgamma}^\mathrm{accr}$ that are free of veiling and photospheric absorption.
\begin{table*}
\centering{
  \caption[Equivalent widths of Brackett-gamma in dwarfs]{Equivalent widths of \brgamma in dwarfs\label{tab:ChaI:brgammaabsorption}}
  \begin{tabular}{cr@{$\pm$}lccr@{$\pm$}lccr@{$\pm$}lccr@{$\pm$}l}
    \hline\hline
    &
    \multicolumn{2}{c}{$W_\mathrm{Br\upgamma}^\mathrm{photosph}$} &
    &
    &
    \multicolumn{2}{c}{$W_\mathrm{Br\upgamma}^\mathrm{photosph}$} &
    &
    &
    \multicolumn{2}{c}{$W_\mathrm{Br\upgamma}^\mathrm{photosph}$} &
    &
    &
    \multicolumn{2}{c}{$W_\mathrm{Br\upgamma}^\mathrm{photosph}$} \\
    SpT &
    \multicolumn{2}{c}{[\AA]} &
    &
    SpT &
    \multicolumn{2}{c}{[\AA]} &
    &
    SpT &
    \multicolumn{2}{c}{[\AA]} &
    &
    SpT &
    \multicolumn{2}{c}{[\AA]} \\
    \hline
    G0 & $-$4.00 & 0.23 &\rule{0.85em}{0cm}& G5 & $-$2.74 & 0.15 &\rule{0.85em}{0cm}& K0 & $-$1.63 & 0.09 &\rule{0.85em}{0cm}& K5 & $-$0.67 & 0.06 \\
    G1 & $-$3.74 & 0.21 &                  & G6 & $-$2.51 & 0.14 &                  & K1 & $-$1.42 & 0.09 &                  & K6 & $-$0.49 & 0.05 \\
    G2 & $-$3.48 & 0.19 &                  & G7 & $-$2.28 & 0.13 &                  & K2 & $-$1.22 & 0.08 &                  & K7 & $-$0.32 & 0.05 \\
    G3 & $-$3.23 & 0.18 &                  & G8 & $-$2.05 & 0.12 &                  & K3 & $-$1.03 & 0.07 &                  & M0 & $-$0.16 & 0.05 \\
    G4 & $-$2.98 & 0.17 &                  & G9 & $-$1.84 & 0.10 &                  & K4 & $-$0.85 & 0.06 &                  & M1 &    0.00 & 0.05 \\
\end{tabular}}
\end{table*}
We assumed that the measured equivalent widths are a good indicator for the accretion, because \brgamma has been found by \citet{ant11} to be well correlated with the accretion luminosity of young low-mass stars, better than the other examined tracers \halpha, \ion{Ca}{II}, [\ion{O}{I}], and Pa$\upbeta$. When the uncertainty of the equivalent width measurement was larger than the derived accretion signal, upper limits were derived from a noise measurement in the continuum surrounding $\lambda_\mathrm{Br\upgamma}$. The resulting fully corrected equivalent widths or upper limits are listed in Table~\ref{tab:ChaI:spectroscopy}.

To infer the presence of an accretion disk around all components, the \emph{disk probability} \citep{dae12a} was calculated for each target component from the measured $W_\mathrm{Br\upgamma}^\mathrm{accr}$, its uncertainty, and the local noise. The values are listed in Table~\ref{tab:ChaI:spectroscopy}.

\subsubsection[Line luminosities and mass accretion rates]{Line luminosities and mass accretion rates: $L_\mathrm{Br\upgamma}$, $L_\mathrm{acc}$, \& $\dot{M}_\mathrm{acc}$}
Line luminosities were derived from the flux-calibrated and dereddened spectra and were converted to accretion luminosities using the emprical correlation $\log(L_\mathrm{acc}) = (1.26\pm0.19)\log(L_{\mathrm{Br}\gamma}/L_\odot)+(4.43\pm0.79)$ from \citet{muz98}. Mass accretion rates $\dot{M}_\mathrm{acc}$ were calculated for all targets with significant accretion according to
\begin{equation}\label{eq:ONC:Mdot}
  \dot{M}_\mathrm{acc} = \frac{L_\mathrm{acc}\,R_*}{GM_*}\left(\frac{R_\mathrm{in}}{R_\mathrm{in}-R_*}\right)
\end{equation}
\citep{gul98}, assuming $R_\mathrm{in}\!\approx\!5R_*$, with stellar radius $R_*$ derived from Stefan's law using the individual stellar luminosities and effective temperatures. Upper limits for target components with no detected \brgamma emission were determined from the minimum-emission feature detectable with 3$\sigma$ significance over the noise in the vicinity of the line. All inferred star and disk parameters are listed in Table~\ref{tab:ChaI:spectroscopy}.

\subsection{New companion candidate to T\,33\,B\label{sec:ChaI:T33}}
T\,33, also known as Glass\,I \citep{gla79}, was discovered to have a binary companion by \citet{zin88}. Our new observations show for the first time clear indications that the secondary component T\,33\,B, which was reobserved by \citet{ghe97} with a separation of 2\farcs5$\pm$0\farcs5 and PA=(284$\pm$5)$^\circ$, itself consists of two stellar components (in the following called T\,33\,Ba and T\,33\,Bb; see Fig.~\ref{fig:obs:ChaIT26T31}). To test whether the newly detected tertiary component is a real companion, we considered component colors, the surrounding stellar density, and proper motion of the components.

The dereddened colors of the two components of T\,33\,B agree roughly with PMS colors \citep{luh10} of K7 and M1 spectral type. Alternatively, a spectral type of K4 can be assigned to the two components assuming that they are giants \citep{bes88}. The spectral type of the combined light was derived to be M0.5$^{+1}_{-1.5}$. Because the total light from the T\,33\,B system is dominated by its brighter component, this confirms the PMS nature of the primary of the T\,33\,B subsystem, since a spectral type as early as K4 can be excluded by the detected photospheric features in \HplusK.

\citet{laf08a} evaluated the probability of chance alignments of multiple components based on the density of stars in the 2MASS catalog in the closest 15\arcmin\ around each star of their sample. For all but the widest components they calculated a probability of $<$\,10$^{-4}$ for chance alignments. Since this includes the 2\farcs4 wide T\,33\,A+B system, the new component located only 0\farcs1 from T\,33\,B must have a considerably lower probability for a chance alignment, making a physical association very likely.

\begin{figure}
  \centering
  \includegraphics[angle=0,width=1.0\columnwidth]{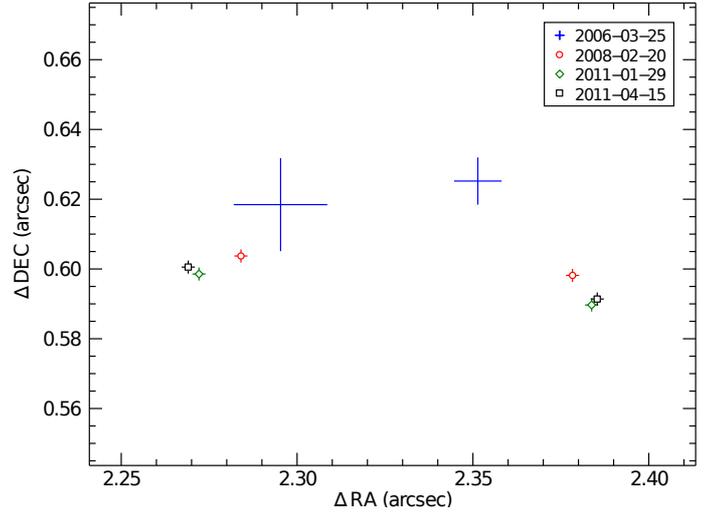}
  \caption[Relative motion of T~33~Ba and Bb]{Position of T\,33\,Ba (right group of points) and T\,33\,Bb (left group) in RA and DEC relative to T\,33\,A at four epochs. The position of T\,33\,Ba \& Bb at the earliest epoch has a comparably large uncertainty since both components are spatially barely resolved.}
  \label{fig:ChaI:T33BaBb}
\end{figure}

To fully ascertain whether Bb is a background object or physically bound to the T\,33 system we used proper motion measurements of the individual components. Two additional epochs of NACO observations taken during the past six years were available in the ESO archive\footnote{2006-03-25 [076.C-0579(A)], 2008-02-20 [080.C-0424(A)], \url{http://archive.eso.org/}}, both of which resolve T\,33\,B into two components. They use a similar instrumental setup as the current observations (\Ks filter, S13 camera). Together with the proper motion of T\,33\,A of $\delta_\mathrm{RA}=-40$\,mas/yr and $\delta_\mathrm{DEC}=+33$\,mas/yr \citep{tei00}, the common proper motion is clearly confirmed: within the $\sim$five years between the first and last of the observed epochs, both T\,33\,Ba and T\,33\,Bb move less than 0\farcs05 in RA and DEC and in different directions relative to T\,33\,A (Fig.~\ref{fig:ChaI:T33BaBb}).
The expected motion of T\,33\,A of $\delta_\mathrm{RA}$=0\farcs20 and $\delta_\mathrm{DEC}$=0\farcs17 in the same time span renders the two components comoving with T\,33\,A, and thus T\,33 is very likely a physically bound triple stellar system.

The residual motion between the components T\,33\,Ba and Bb is probably orbital motion. An estimate of the shortest orbital period using Newton's and Kepler's laws under the assumption of a circular orbit and that the last observation was at maximum elongation returns $P_\mathrm{min}$=102$\pm$15\,yrs. Classification of T\,33\,BaBb as a binary that has been hiding its binary nature through aligning its components in an inclined orbit during the past few years explains why its binary nature has not been discovered before and why \citet{ngu12} observed a deviant radial velocity of T\,33\,B from the rest of the cluster while no signal of spectroscopic binarity was detected.

\section{Discussion}
\subsection{Stellar parameters and sample biases}
\subsubsection[The extended and restricted samples]{\emph{Full}, \emph{extended}, and \emph{restricted} samples}\label{sec:ChaI:restrictedsample}
\emph{The full sample} of 26 multiple (=\,52 individually detected) objects comprises targets from a broad range of binary separations, component spectral types, and observational quality. To ensure the comparability and validity of the results, we defined in addition to the \emph{full sample} of components with individually obtained spectroscopy (irrespective of the fact that they might be close binaries) two new samples, the \emph{extended sample} and the \emph{restricted sample}.

The full sample consists of all components ``as observed''. This disregards that some of these components may be unresolved visual or spectroscopic binaries. The \emph{extended sample}, in contrast, is composed of all known individual components. Spectroscopic binaries as well as known visual components that remain unresolved by the present study each enter the sample as two individual components. Since not all individual stellar parameters are known, this sample will only be used for number statistics such as the disk frequency (Sect.~\ref{sec:ChaI:diskfrequency}), where the presence of disks in unresolved binaries can be inferred from the unresolved data. This sample grants the most pristine view to the data, since no additional restrictions to the original component sample are applied. The extended sample consists of 62 components (19 binaries, 4 triples, and 3 quadruple stars).

\emph{The restricted sample} was composed to eliminate observational and compositional biases by applying the following selection criteria. (a) Spectral types must be in the range for T\,Tauri stars, i.e., later than G0. (b) All target components must be stellar. It is not clear whether the formation process of brown dwarfs is identical to that of stars. To exclude the effects of possibly different initial conditions, all brown dwarf candidates in the sample were excluded. (c) The information must be attributable to one stellar component. A few stars in the sample could not be resolved with our adaptive optics spectroscopy or are suspected spectroscopic binary candidates. This requirement limits the restricted sample to binaries with projected separations of $\gtrsim$\,0\farcs15 or $\gtrsim$\,25\,au at the distance of Cha\,I. 
While these restrictions mean that some of the binary components are ``singled'', i.e., only one component enters the sample, it reduces the slight preference for primaries to have earlier spectral types than secondaries (see next paragraph and Fig.~\ref{fig:ChaI:SpTs}) since early-type primaries and late-type secondaries are removed.

The restricted sample comprises 43 components. If \emph{binary} parameters are to be investigated with respect to the restricted sample, only binaries with both components in the restricted sample are considered. This \emph{restricted binary sample} consists of 17 binaries.

\subsubsection{Spectral types}
\begin{figure}
  \centering
  \includegraphics[angle=0,width=1.0\columnwidth]{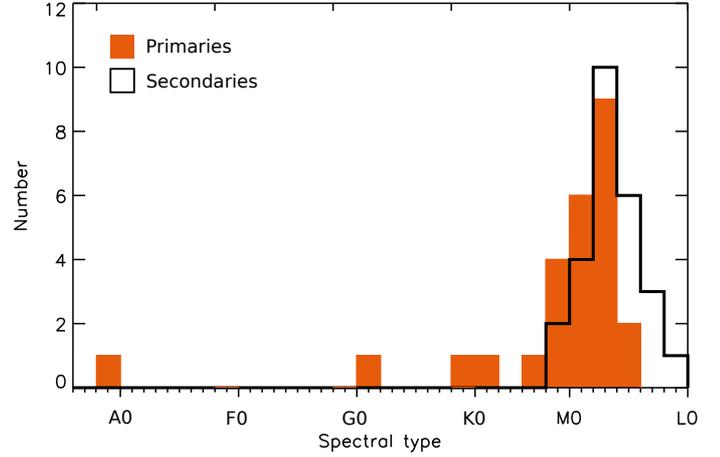}
  \caption[Spectral types histogram]{Histogram of spectral types divided into primary stars (solid orange) and secondary stars (black outline). The spectral types of secondaries are on average 1.5 subclasses later. Unresolved close binaries in triple systems are plotted as one component.}
  \label{fig:ChaI:SpTs}
\end{figure}

As can be seen in the spectral type distribution in Fig.~\ref{fig:ChaI:SpTs}, the median spectral type of primary stars (M1.5) is earlier than that of the secondaries (M3).
This is true for the full and restricted samples. To test whether the subsample of binary stars is representative for the population of Cha\,I members, we calculated the probabilities that the spectral types of primaries and secondaries are drawn from a common parent distribution with single stars of Cha\,I \citep{luh04,luh07}, respectively. We limited the single-star sample to V$<$16.7\,mag to account for the fact that the primaries were selected to serve as SINFONI adaptive optics reference stars. The resulting KS probability of 92\% indicates that the distribution of primary star spectral types in this sample cannot be distinguished from the single star distribution. Secondary stars, however, are not subject to this selection bias and agree with the unconstrained sample of single stars (KS probability 19\%).

The apparent difference between the primary and secondary spectral type distributions in Fig.~\ref{fig:ChaI:SpTs} by definition reflects the mass differences between the sets of primary and secondary components. This can produce biases when comparing the parameters of both populations since many parameters, such as stellar evolutionary timescales, are mass-dependent with the more massive components typically evolving faster than the less massive stars. The resulting differences of the inferred parameters are discussed in the subsequent sections when they are suspected to occur.

\subsubsection{Relative extinctions and ages\label{sec:ChaI:relativeages}}
\begin{figure}
  \centering
  \includegraphics[angle=0,width=0.8\columnwidth]{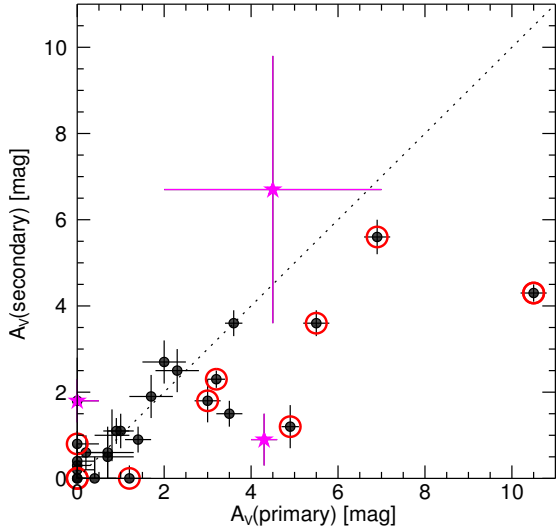}
  \caption[Primary versus secondary extinction]{Primary vs.\ secondary extinctions $A_V^\mathrm{phot}$. Target binaries with at least one component with strong near-IR color excess E(\KsLp)\,$>$\,0.5 are marked by a red circle. The dotted line shows equal extinctions in both compononents. The three magenta stars (CHXR\,73, T\,26, T\,33) have at least one component without measured E(\KsLp). From their very red color of \HK\,$>$0.6\,mag a strong near-IR excess can also be inferred for these binaries.}
  \label{fig:ChaI:AVprimary_vs_AVsecondary}
\end{figure}

In Fig.~\ref{fig:ChaI:AVprimary_vs_AVsecondary} we compare the photometric extinctions in both components of each binary.
While most of the extinctions are identical for the components of the same binary, the multiples with significant differences in component extinction are mostly host to at least one very red component with near-infrared excess E(\KsLp)\,$>$\,0.5. If extinction is dominated by the degree of embeddedness of a binary in the molecular cloud, it is expected to be similar for both components of a binary. The presence of \KsLp color excess apparently leads to a different photometric extinction value for the respective component. Since the spectroscopically determined extinctions (see Table~\ref{tab:ChaI:spectroscopy}) show no such correlation (the values scatter almost uniformly around equal primary and secondary extinctions, though with a larger spread due to the larger uncertainties of $A_V^\mathrm{spec}$), additional systematic uncertainty of the photometric extinctions as a function of color excess may be present that do not affect the spectroscopically derived extinctions. A possible origin can either be an overestimation of the slope of the extinction vector in the (\JH)--(\HKs) color-color diagram (Fig.~\ref{fig:ChaI:colorcolordiagram}) or an underestimation of the slope of the CTTS locus (which can be caused, for example, by assuming filter system transformations that do not accurately describe the filters in use) the result would be a mostly positive correlation of $A_V$ with the color-excess, as observed. In the following, we used the derived extinctions as measured and caution that a possible systematic uncertainty may be present with a median amplitude of 2\,mag for high E(\KsLp) components. If true, the luminosity of the respective components would be overestimated by on average 0.21\,dex, comparable in size to the random uncertainty of the luminosity values, resulting in an age estimate that is younger than those derived in Table~\ref{tab:ChaI:spectroscopy} by $\Delta\log\tau\!\approx\!0.3$ at M\,=\,0.4\,M$_\odot$.

\begin{figure}
  \centering
  \includegraphics[angle=0,width=0.8\columnwidth]{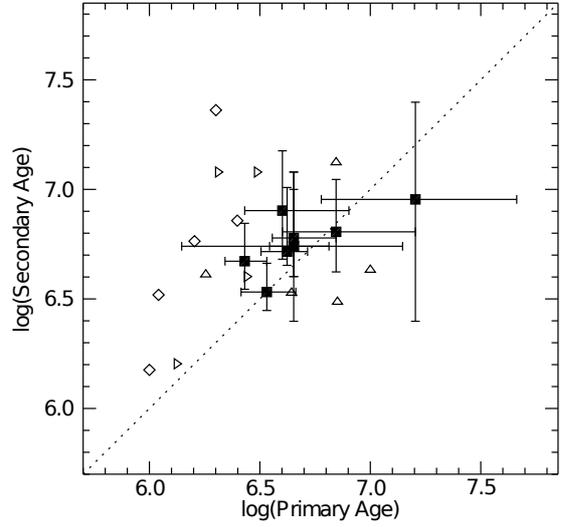}
  \caption[Primary age versus secondary age]{Primary age vs.\ secondary age in logarithmic units. Open symbols are target binaries with unreliable age measurements because either their primary (triangles right) or secondary (triangles up) or both components (diamonds) are spectroscopic binaries or exhibit a strong color-excess E(\KsLp)\,$>$\,0.5 (or \HKs$\!>$\,0.6, if \KsLp could not be measured). Filled symbols with error bars are binaries with spatially separated components with low color excess. The dotted line is the reference for coeval ages.} 
  \label{fig:ChaI:PrimaryAge_vs_SecondaryAge}
\end{figure}

Taking these effects into consideration, the measured ages show that binary components are coeval. Fig.~\ref{fig:ChaI:PrimaryAge_vs_SecondaryAge} compares the logarithmic isochronal primary and secondary ages from Table~\ref{tab:ChaI:spectroscopy}.
Binaries with at least one component with strong color excess or an unresolved binary component are mostly located off the sequence of equal-ages. This is due to overestimated luminosities of components with strong continuum excess or overestimated extinction and accordingly under-estimated ages. In contrast, all binaries with low-excess components are consistent with equal age components to within 1$\sigma$. To test whether the observations are compatible with the simultaneous formation of the two components of a binary, we compared the logarithmic age differences $|\Delta\log\tau|=|\log\tau_\mathrm{prim}-\log\tau_\mathrm{sec}|$ following \citet{kra09}. The rms scatter in $|\Delta\log\tau|$ is 0.03\,dex, which is compared with 10000 randomly paired samples of the same size as the target sample, drawn from the primaries and secondaries with low veiling. Only 2\% of the random samples have an rms scatter smaller than 0.03\,dex. The binary sample is thus more coeval than the total sample of binary components with a significance of 2.3$\sigma$.

The median inferred age of the subset of low-$r_K$ components (excluding unresolved binaries) is 5.5\,Myr, older than the 2--3\,Myr derived by \citet{luh07} for a large sample of Cha\,I members with models of \citet{bar98} and \citet{cha00}. Even the systematically younger ages when using the \citet{pal99} models (see Fig.~\ref{fig:ChaI:HRdiagramPalla}) produce a median age (4\,Myr) that is older than the single-star value. Since it is unlikely that binaries are intrinsically older than single stars of the same region, this discrepancy must be due to either sample biases or a systematic difference in the derivations of luminosities and effective temperatures. While part of the difference might arise from the isochrones used, we found that the luminosities derived by \citet{luh07} are systematically higher than those derived in the binary sample. Part of this difference is due to the larger distance that \citeauthor{luh07} assumes. The increase of the distance from 160 to 168\,pc requires a luminosity increase of $\Delta\log L$\,=\,0.05 for all targets to match the observed \J-band magnitudes. Furthermore, while a correction factor for unresolved binarity was applied by \citet{luh07}, he did not account for possible excess emission in \J-band, which can lead to comparably high luminosities, as we see for the high-veiling components in the HR-diagram (Fig.~\ref{fig:ChaI:HRdiagram}). This can lead to an underestimation of the median age. In fact, if one includes all binary components of the present sample irrespective of their veiling into the age estimation, a median age of 4.4\,Myr is derived from the \citet{sie00} models and 2.6\,Myr from \citet{pal99}.

A thorough absolute age determination should be repeated with a larger sample than the one presented here. Nevertheless, the age determination serves to confirm the similar \emph{relative} ages of the binary components as discussed above.
Coevality and equal component extinctions are in line with previous findings of binaries in Cha\,I \citep{bra97} and other star-forming regions \citep{whi01,kra09}.

\subsection[Evolution of disks in Cha~I binary components]{The evolution of disks in Cha\,I binary components}
\subsubsection{Disk frequency\label{sec:ChaI:diskfrequency}}
From the number of target components with \KsLp color excess that were used as an indicator for the presence of hot dust in the inner disk and those with significant \brgamma emission, which indicate ongoing accretion activity, we determined the frequency of dust and accretion disks around the components of binaries in Cha\,I and compared it with disks around single stars of this region.

\paragraph{Accretion disks.}
The fraction of target components harboring accretion disks was determined from the individual disk probabilities in Table~\ref{tab:ChaI:spectroscopy} as described in the appendix of \citet{dae12a}.
Some components have no good estimates for the veiling level. A correction of \Wbrgamma is not possible for these targets, and the accretion strength might be significantly larger than measured. One of the reasons for an inaccessible veiling value is high veiling itself, since the line features used for the veiling estimation can be reduced in depth until they blend into the continuum noise. Excluding these targets may accordingly bias the sample against accretors. To avoid this bias, these targets were included with a neutral disk probability of 0.5. As a test, we also derived disk fractions assuming the limiting cases of 0\% and 100\% disk probability and found no significantly different results. In the following, we consider the extended and the restricted samples.

The extended sample of all components returns an accretion disk fraction of $F^\mathrm{ext}=18^{+8}_{-6}$\%. This calculation is possible since all unresolved binaries except T\,31\,A show no \brgamma emission and can be treated as consisting of two non-accreting components. T\,31\,A is a suspected SB2 and shows signs of accretion (\Wbrgamma\,$>$\,9.4\,\AA). The uncertainty of being host to one or two accreting components is included in the calculation of the disk frequency. 

The restricted sample returns a very similar frequency of $F^\mathrm{restr}=16^{+9}_{-6}$\%. Both calculations include only the information from the \brgamma emission measurements. It is known that \brgamma returns a lower fraction of accretors than the \halpha feature, which, however, defines the classes of classical and weak-line T\,Tauri stars. This is due to the $\sim$100 times lower transition probability of \brgamma excitation compared to \halpha\footnote{\emph{NIST} Atomic Spectra Database, \url{http://www.nist.gov/pml/data/asd.cfm}\,.}. With the help of the correction factor $f=0.125$ derived in \citet{dae12a} from the fraction of accretors detected in \halpha and \brgamma and a number of $N$\,=\,43 components in the restricted sample, the number of CTTS without \brgamma emission is estimated to be $f/(1-f)\times (F\cdot N)\approx\!1$. This increases the restricted accretion disk fraction to $F^\mathrm{restr}_\mathrm{corr}=19^{+9}_{-6}$\%.

This fraction is significantly lower than the accretion disk fraction of single stars in Cha\,I of $F^\mathrm{single}=44\pm8\%$ \citep{moh05}. Assuming that binaries and single stars are born at the same time in the star formation history of a cluster, this indicates that binaries with separations of $\sim$25--1000\,au finish accreting on a shorter timescale than single stars. To investigate this trend in more detail, we divided the sample into binaries with separations closer and wider than 100\,au, close to the median separation of 119\,au of the sample. The resulting fractions of $F^\mathrm{restr}_\mathrm{<100au} = 10^{+15}_{-5}\%$ and $F^\mathrm{restr}_\mathrm{>100au} = 23^{+11}_{-8}\%$ show that the accretion disk fraction is particularly low in the closest binaries of the sample, less than half of the wider sample. This corroborates theoretical expectations that the timescale for disk evolution is significantly shortened in binaries with separations $<$100\,au, if one assumes a correlation of disk lifetime with the size of the (truncated) disk \citep[e.g.,][]{mon07}. The fact that the reduction of disk frequency in wide binaries is less pronounced but still significant indicates that disk lifetimes can be shortened even if the truncation of the outer disk is only weak. This agrees with the results from the ONC \citep{dae12a}, where the sample of binaries does show a reduction of accretion disk frequency although only separations of $\sim$100--800\,au and no closer pairs could be examined owing to the comparatively large distance of the ONC.

\begin{figure}
  \centering
  \includegraphics[angle=0,width=0.8\columnwidth]{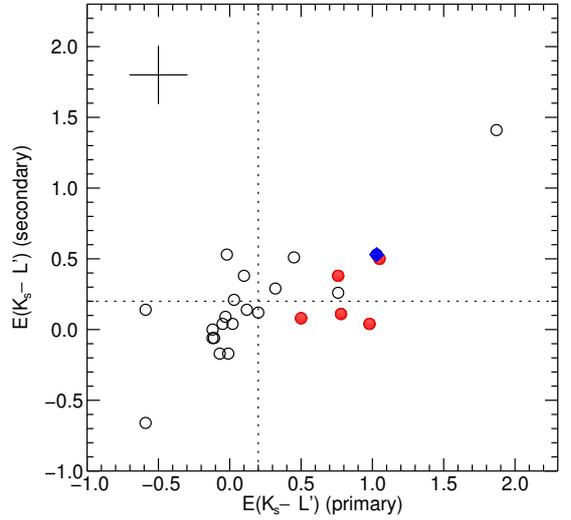}
  \caption[Near-infrared excess E(Ks-L') of primary and secondary components]{Near-infrared excess \EKsLp for the primary and secondary components of each binary with individual component photometry. Red filled circles and blue filled diamonds are binaries with accreting primaries and secondaries. Targets to the right and above the dotted lines have significant near-IR excess, indicating the presence of inner disk material. A typical error bar, dominated by the systematic uncertainty of $\sim$0.2\,mag, is indicated in the upper left.} 
  \label{fig:ChaI:EKLprimary_vs_EKLsecondary}
\end{figure}

\paragraph{Inner dust disks.}
The frequency of inner dust disks $\mathfrak{F}$ is measured from the number of targets with significant \KsLp excess in the nearly simultaneous photometry. A diagram of primary and secondary \EKsLp is shown in Fig.~\ref{fig:ChaI:EKLprimary_vs_EKLsecondary}.
As derived in Sect.~\ref{sec:ChaI:colorcolor}, the systematic uncertainty of the \KsLp excess measurements is about 0.2\,mag. Accordingly, \KsLp emission is in significant excess over the pure stellar color for components with \EKsLp$>0.2$. There are 21$^{+17}_{-7}$ components out of 55 with measured \KsLp colors in the extended sample fulfill this criterion, and 20$^{+12}_{-6}$ components in the restricted sample of 40 components with measured colors. Uncertainties of the number counts were determined from targets for which \EKsLp$=0.2$ is within the uncertainty limits. The resulting inner dust disk fractions are $\mathfrak{F}^\mathrm{ext}=38^{+31}_{-13}$\,\% and $\mathfrak{F}^\mathrm{restr}=50^{+30}_{-15}$\,\%. 

The values agree well with previous measurements of the disk frequency in binaries of Cha\,I. \citet{dam07} measured the frequency of targets with 8 or 24\,\mum excess in 15 \emph{spatially unresolved} binaries and 2 triples in Cha\,I and derived a nominal frequency of multiples with disks of $\mathfrak{F}=35^{+15}_{-13}$\%.
Their value, obtained from 10--300\,au binaries, agrees well with the fraction of components with \KsLp excess obtained from the extended sample. And indeed, when applying the selection criteria of the restricted sample to the \citeauthor{dam07} binary sample, the inferred hot dust disk fraction increases to 42\%, since the fraction of binaries with separations $\lesssim$\,25\,au that show no signs of dust excess is high (four out of five). This is compatible with the fraction found for the restricted sample of binaries. 

The extended and in particular the restricted inner disk fractions of Cha\,I binaries are compatible with the average Cha\,I single star value of $\sim$47--55\% (with weak fluctuations depending on stellar mass) found from photometric excess in the \emph{Spitzer}/\emph{IRAC} bands between 3.6\,\mum and 24\,\mum \citep{luh08b,dam07}. A significant reduction of disk presence compared to single stars is mostly visible in close binaries: out of 16 components that are part of spectroscopic or visual binaries with separations $<$\,25\,au, only 3 or 4, depending on whether the spectroscopic binary T\,31\,A has one or two disk-bearing components, were found with significant \KsLp excess, a fraction of 19--25\%. The inclusion of close binaries explains why the extended sample shows a lower disk fraction than the restricted sample.

\begin{figure}
  \centering
  \includegraphics[angle=0,width=1.0\columnwidth]{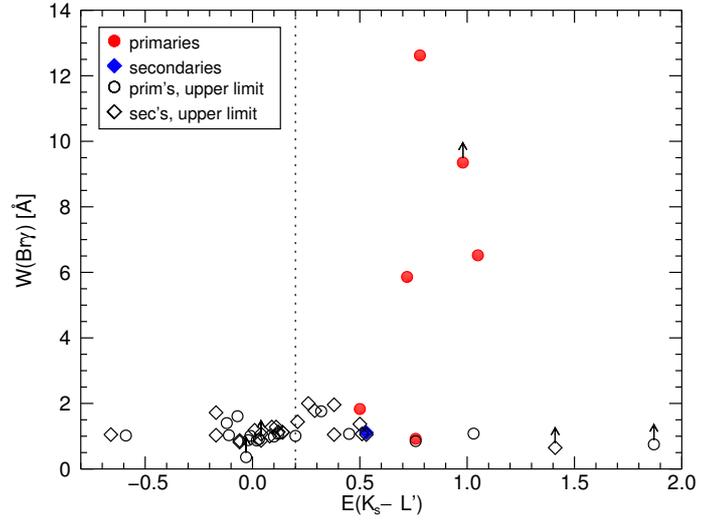}
  \caption[Brackett-gamma emission vs.\ near-infrared color excess E(Ks-L')]{\brgamma equivalent width as a function of near-IR color excess \EKsLp. Filled symbols show accretion detections, open symbols are upper limits of \Wbrgamma. Primaries and secondaries are denoted by circles and diamonds. Stars to the right of the dotted line have significant \KsLp excess. Targets without veiling measurement can have larger \Wbrgamma than shown here since the correction factor $(1+r_K)$ could not be applied; they are marked with arrows.}
  \label{fig:ChaI:brgammavsEKL}
\end{figure}

\paragraph{The correlation of inner disk presence and accretion.}
The fraction of inner dust disks around binary components is significantly higher than the accretion disk fraction. Figure~\ref{fig:ChaI:brgammavsEKL} compares measured \brgamma equivalent widths with the near-IR color excess.
While all accreting components show significant color excess, a nominal fraction of 65\% of the binary components with near-infrared excess shows no signs of accretion in \brgamma. This reduces to 55\% when accounting for the fact that ISO\,126, the binary with the two components with the strongest near-IR excess, has been found to be accreting in \halpha studies \citep{saf03}. This fraction is significantly higher than for Cha\,I single stars according to the study by \citet{dam07}, who found 6 of 21 stars with inner dust disks (inferred from [4.5]$-$[8.0] excess) without signs of accretion (\halpha).

This discrepancy of the accretion and dust disk presence has been observed not only for Cha\,I, but in several star-forming regions, making a stochastic explanation accretion variability unlikely. We conclude that the approximately exponential decay of disk frequency is more rapid for accretion disks (time constant $\tau_\mathrm{acc}\approx2.3$\,Myr) than for hot inner dust disks \citep[$\tau_\mathrm{dust}\approx3$\,Myr;][]{fed10}. Binary companions in Cha\,I apparently amplify the effect that accretion disks disappear before dust disks around the same stars. Possible reasons are an inside-out disk evolution, UV irradiation from the star, or planet formation and migration. A detailed discussion will follow in a forthcoming paper (Daemgen et al.\ 2013).

\subsubsection{Differential disk evolution\label{sec:ChaI:differentialdiskevolution}}
\begin{figure}
  \centering
  \includegraphics[angle=0,width=1.0\columnwidth]{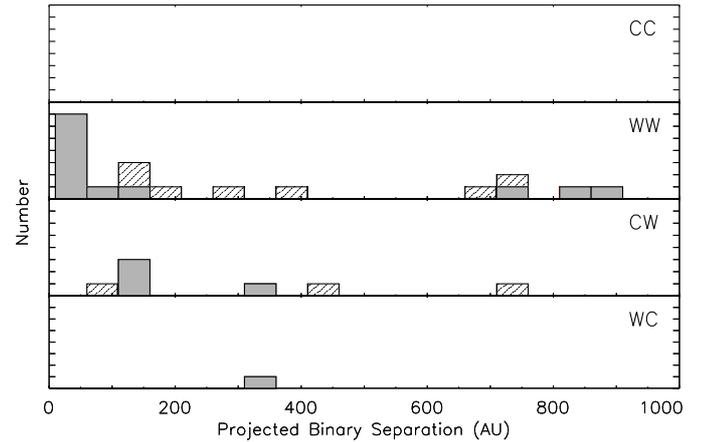}
  \caption[Accretion status as a function of binary separation]{Accretion status as a function of binary separation. The gray shaded bars show the restricted sample, the combined shaded and hatched bars represent the full sample. One tick mark on the y-axis is equivalent to one binary. The classifications CC, WW, CW, and WC refer to binaries composed of {\bf c}lassical and {\bf w}eak-line T\,Tauri stars with the first and second character describing the primary and secondary component. No systems with two accreting components were detected.}
  \label{fig:ChaI:CCWWCWWC_vs_separation_histogram}
\end{figure}

Fig.~\ref{fig:ChaI:CCWWCWWC_vs_separation_histogram} shows the separation of Cha\,I binaries classified by their accretion status as CC (both components accreting), WW (no accreting components), CW and WC (mixed pairs with the primary and secondary component accreting, respectively). 
While the majority of binaries in the restricted sample were found to have no accreting components ($12^{+3}_{-1}$/17\,$=$\,70$^{+18}_{-11}$\% WW), there are no CC binaries ($0^{+1}_{-0}$) detected and all accreting components are part of mixed systems with either an accreting primary ($4^{+3}_{-2}$$\times$CW) or secondary ($1^{+0}_{-1}$$\times$WC). A qualitatively identical picture is posed by the full sample. The population of these classes agrees well with random pairing of the number of individual accretors and non-accretors in the restricted sample. About one CC binary would be expected, $\sim$four mixed pairs, and the rest ($\sim$12) are expected to be WW. This indicates that the existence of a disk around one Cha\,I binary component has no impact on the probability of finding another disk around the other component, and both components can be treated as independent in terms of their disk probability. The effect of disk synchronization that was detected for ONC binaries \citep{dae12a} is not observed here. This is probably because Cha\,I is older than the ONC. However, a preference for mixed systems with an accreting primary can be observed. WC systems appear more scarce than CW as observed in previous studies of Taurus and other star-forming regions \citep{mon07}, but it was not confirmed by, e.g., data of the ONC \citep[4$\times$CW and 3$\times$WC;][]{dae12a}.

The paucity of WC systems in Cha\,I points to either a bias against their detection or a different formation and/or evolution of secondary disks as a function of the cluster environment. Since binaries in this sample were selected from an adaptive optics search for companions, no preference for or against accreting secondary components is expected. The accretion measurements might even be in favor of the detection of accretion in the fainter secondary spectra since the detection limit of the mass accretion rate is lower the fainter the system. In fact, the only accreting secondary component in this study has the lowest detected mass accretion rate, below the upper limits of most other components (see Sect.~\ref{sec:ChaI:Mdot} and Fig.~\ref{fig:ChaI:Mdot_vs_M}). The fact that, independently of spectral type, secondaries systematically show no or only very weak accretion indicates that we see a real effect of the binary environment rather than a selection effect. A real paucity of mixed systems with accreting secondaries in Cha\,I is therefore the most plausible explanation. Assuming that almost all binaries start in a CC configuration \citep[more than 60\% of all binaries in the very young 1--2\,Myr Taurus association are CC][]{whi01,mcc06,ken95}, this observation points to a more rapid evolution of the disk around the less massive component. 

\begin{figure}
  \centering
  \includegraphics[angle=0,width=1.0\columnwidth]{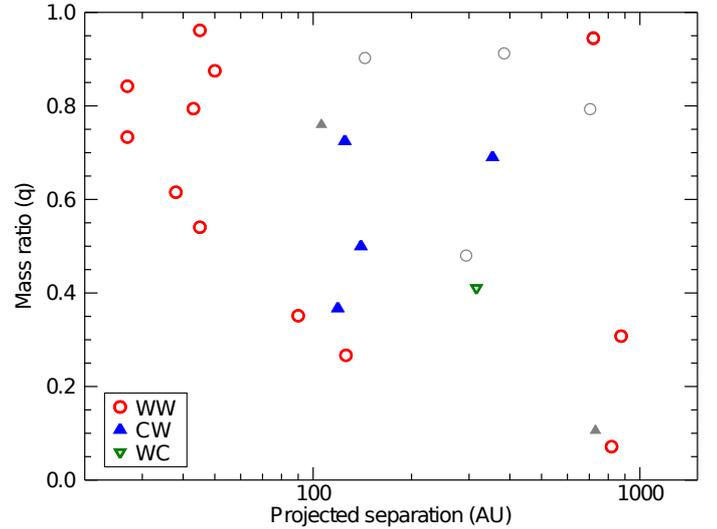}
  \caption[Mass ratio as a function of binary separation]{Mass ratio $q=M_\mathrm{sec}/M_\mathrm{prim}$ as a function of binary separation. Red open circles show binaries with two non-accreting components (WW), blue triangles (up) show mixed systems with an accreting primary and green triangles (down) with an accreting secondary from the restricted sample. Gray symbols represent targets that are in the full sample but not in the restricted sample. There are two binaries (Hn\,4, T\,39\,AB\footnotemark\ both WW) where the mass of the fainter component is calculated to be higher than the primary's mass. The inverse $q^{-1}$ is plotted for these.}
  \label{fig:ChaI:massratio_vs_separation}
\end{figure}

While the above considerations show that the spectral type and thus the absolute mass of the binary components is unlikely to cause a preference for mixed systems, we explored the possibility that the relative disk lifetimes of primaries and secondaries are governed by the \emph{relative masses} of the stellar components, i.e., the component mass ratios $q=M_\mathrm{sec}/M_\mathrm{prim}$. Fig.~\ref{fig:ChaI:massratio_vs_separation} shows the mass ratio as a function of binary separation, indicating the accretion type as well.
The diagram features mixed systems only with mass ratios below $q\!=\!0.8$, while 5 out of 12 WW systems measure $q\!\gtrsim\!0.8$. This suggests a mass ratio-dependent mechanism: owing to its smaller Roche lobe size, disk truncation is stronger for the less massive component, which may lead to a shorter disk lifetime due to an early-ceasing angular momentum transport \citep{art94}. A reduction of secondary disk lifetime agrees with theoretical expectations (it scales with disk radius $R$ roughly as $R^{-0.5..1}$ for a viscous disk), although a significant difference between primary and secondary disk lifetime is only expected for mass ratios $q\!<\!0.5$ \citep{mon07}.
\footnotetext{T\,39\,B is a spectroscopic binary candidate, which might explain its higher inferred mass.}

While mass ratios together with viscous disk evolution can explain at least part of the observed CW systems, the more massive component is always predicted to keep its disk for longer. Accordingly no WC systems would form. That these systems are nevertheless observed might be partly due to projection effects. Disks in very wide systems that only appear close due to projection may evolve without notice of the other star if their outer tidal truncation radius is larger than the size of a typical disk, i.e., for binary separations $\gg$1000\,au. On one hand, unrelated systems can form in any configuration since the lifetime of an individual disk can be anywhere between $<$\,1 and $\sim$10\,Myr. On the other hand, WC systems might even be preferred for very wide and unrelated components, since it was found that the frequency of disks around single stars declines with increasing stellar mass \citep[e.g.,][]{lad06,car06}.

\subsubsection[Mass accretion in Cha~I binaries]{Mass accretion in Cha\,I binaries\label{sec:ChaI:Mdot}}
\begin{figure}
  \centering
  \includegraphics[angle=0,width=1.0\columnwidth]{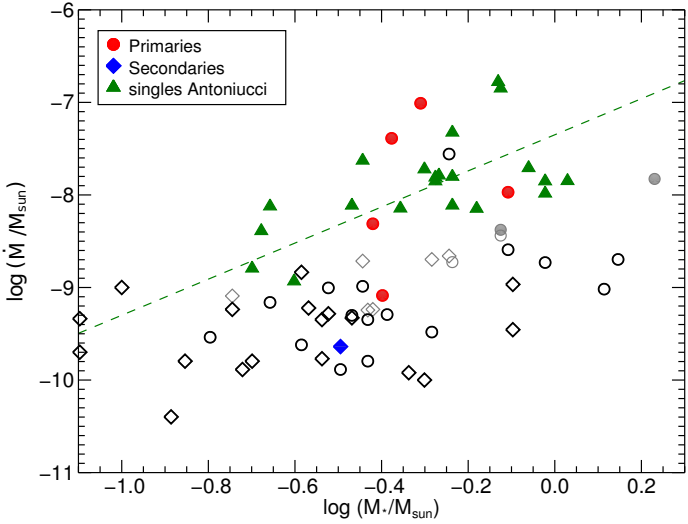}
  \caption[Mass accretion rate as a function of stellar mass]{Mass accretion rate as a function of stellar mass. Filled circles and diamonds are significantly accreting primaries and secondaries of Cha\,I binaries. Open symbols are upper limits of undetected binary components. Colored symbols denote the restricted sample, gray symbols the full sample only. Green filled triangles show Cha\,I single stars also measured from \brgamma by \citet{ant11}. These authors also derived a power-law fit to their data, included as a green dashed line. Upper-limit values of the \citeauthor{ant11} non-detections are not published.}
  \label{fig:ChaI:Mdot_vs_M}
\end{figure}

Mass accretion rates are shown as a function of stellar mass in Fig.~\ref{fig:ChaI:Mdot_vs_M}.
The new Cha\,I binary data are compared with single star mass accretion rates from \citet{ant11}. Because these were obtained from the same indicator, \brgamma emission, no systematic biases are expected. The distributions of the two populations, Cha\,I binary components and single stars, qualitatively agree in the $\log(\dot{M})$--$\log(M_*)$ diagram. There are, however, too few accretors in the binary sample to statistically confirm or reject agreement with the relation $\dot{M}\!\propto\!M_*^{1.95}$ proposed by \citet{ant11}. 

Interestingly, all accreting components in this sample were observed at masses $\ge$\,0.32\,M$_\odot$, i.e., 100\% of all accretors are among the most massive $\sim$60\% of all components in the binary sample. A KS-test returns a 87\% chance that the masses of accretors and nonaccretors are drawn from different parent populations. This is neither due to limited sensitivity -- the detection limits are mostly below the $\log(\dot{M})$--$\log(M_*)$ relation of accretors by about 0.5 magnitudes -- nor a general property of Cha\,I accretors: \citet{ant11} did detect single stars with masses $<$\,0.32\,M$_\odot$\ with significant accretion. This corroborates the previously inferred low relevance of spectral type or mass for the existence of accretion signatures of secondary stars: since low-mass accretors $<$\,0.32\,M$_\odot$\ do exist in Cha\,I singles, but none are observed in binaries (neither primary nor secondary), it is not the lower mass of the secondary that reduces the chance of dealing with an accreting star, but the presence of a more massive stellar companion is the most likely cause for the disappearance of the disk around the secondary.

While these considerations suggest a dependence of accretion parameters on binary parameters, the small number of binaries in Cha\,I and the low fraction of accretors limits the statistical significance of these results in Chamaeleon\,I. Future exploration of mass accretion rates must therefore rely on the compilation of data from various star-forming regions. 

\section{Conclusions}
We have presented an observational near-infrared study of the individual components of 19 T\,Tauri binary stars and 7 triples, including one newly resolved tertiary component, in the Chamaeleon\,I star forming region. The target sample covers close to all currently known multiples in the separation range between 0\farcs2 and 6\arcsec. Imaging data in \JHKsLp bands were taken with VLT/NACO together with close-in-time 1.5--2.5\mum integral field spectroscopy with VLT/SINFONI, covering the accretion-induced Brackett-$\upgamma$ feature. The combined data set enabled us to measure the relative position, magnitude, color, spectral type, and the equivalent width of \brgamma emission of each individual component. Stellar luminosities and effective temperatures were calculated, which enabled us to derive stellar component masses, ages, and radii through a comparison with pre-main sequence evolutionary tracks. This information was combined with the derived probabilities for the existence of hot circumstellar dust, as inferred from \KsLp color excess, and accretion disks from \brgamma emission. The data enable us to conclude the following:
\begin{enumerate}
  \item The components of a binary are more coeval than random pairs of Cha\,I binary components, with a significance of 2.3$\upsigma$.

  \item Components of binary stars in Cha\,I show signs of accretion less often than single stars in the same region. From the new data of T\,Tauri binaries with separations of 25--1000\,au, a fraction of $F=19^{+9}_{-6}$\% of all stellar components has been inferred to harbor an accretion disk. This is less than half of the single star disk fraction in Cha\,I of $F^\mathrm{single}=44\pm8\%$. Apparently, the presence of a stellar binary component causes accretion to cease earlier. A reduction of the accretion disk frequency is particularly pronounced in close binary systems of separations 25--100\,au where the accretion disk frequency $F_\mathrm{<100au}=10^{+15}_{-5}$\% is less than half of the accretion fraction in wider binaries (100--1000\,au, $F_\mathrm{>100au}=23^{+11}_{-8}$\%). 

  \item The dust disk fraction of binary components is indistinguishable from that of single stars. $\mathfrak{F}=50^{+30}_{-15}$\,\% of all binaries with separations between 25 and 1000\,au were found with \KsLp excess, identical to the measured value in singles. The frequency of hot circumstellar dust around the components of binary systems with separations $\lesssim$\,25\,au, in contrast, is reduced to $\lesssim$\,25\%.

  \item All accreting components were found to be members of mixed pairs of accreting and non-accreting components. The presence of a disk around either binary component appears to be unrelated with the disk around the other component, suggesting no synchronized disk evolution. Mixed systems were not found with equal component masses, but their mass ratios are always below $q$\,=\,0.8. Pairs without accreting components occupy a range of mass ratios including close-to equal-mass binaries with $q$\,$>$\,0.8. Unequal component masses and the induced stronger disk truncation of the less massive component are a possible explanation for the differential disk evolution.

  \item Out of the seven identified accretors only one is a secondary star, i.e., the less massive star in the respective binary system. This is also the weakest accretor found in the sample. This suggests that accreting secondaries are rare and/or only weakly accreting. The existence of single stars with masses comparable to those of the binary components but significantly stronger accretion indicate that this is a consequence of the presence of a higher-mass stellar companion to a disk-hosting star.

  \item The measured mass accretion rates of binary components agree well with the mass accretion rates of single stars of Cha\,I for components $>$\,0.3\,M$_\odot$. Accreting binary components with masses below 0.3\,M$_\odot$, however, are less frequently observed than around single stars.
\end{enumerate}

\begin{acknowledgements}
We thank the anonymous referee for a thoughtful report. This research has made use of the SIMBAD database, operated at CDS, Strasbourg, France. It has used data products from the Two Micron All Sky Survey, which is a joint project of the University of Massachusetts and the Infrared Processing and Analysis Center/California Institute of Technology, funded by the National Aeronautics and Space Administration and the National Science Foundation. This publication is supported by the Austrian Science Fund (FWF).
\end{acknowledgements}

\bibliographystyle{aa}
\bibliography{ms}

\appendix
\section{Spectra of the individual target components}\label{sec:app}
\begin{figure*}
  \centering
  \includegraphics[angle=0,width=1.0\textwidth]{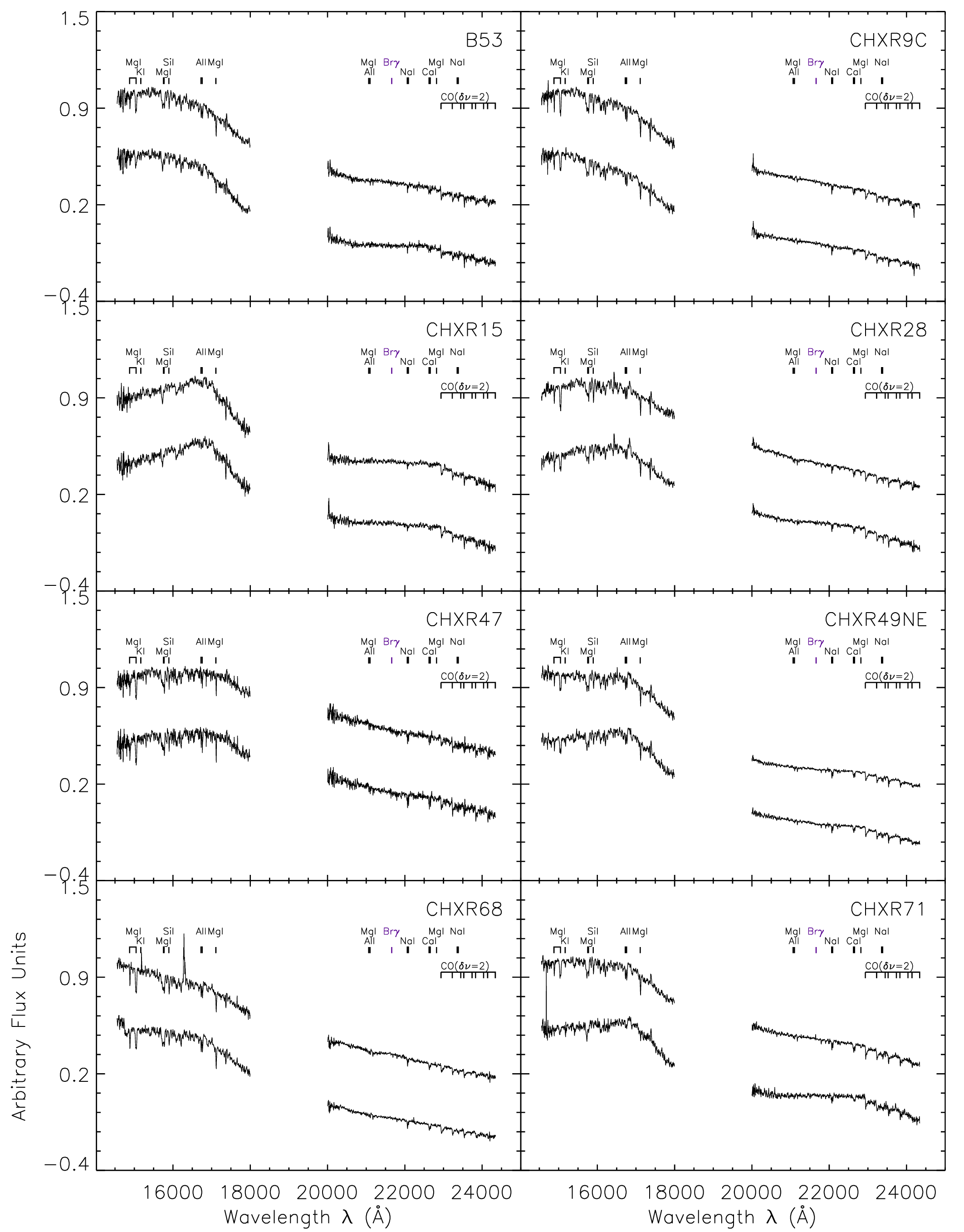}
  \caption[SINFONI spectroscopy of Cha~I binaries]{Spectroscopy of all target components observed with SINFONI. The most massive component is shown at the top, the least massive component at the bottom. The most prominent absorption features of low-mass pre-main sequence stars are indicated (see Table~\ref{tab:obs:features} for a list). The spectra of CHXR\,73\,B and T\,14\,B are shown binned by a factor of 4.}
  \label{fig:obs:ChaIspectra}
\end{figure*}
\addtocounter{figure}{-1}
\begin{figure*}
  \centering
  \includegraphics[angle=0,width=1.0\textwidth]{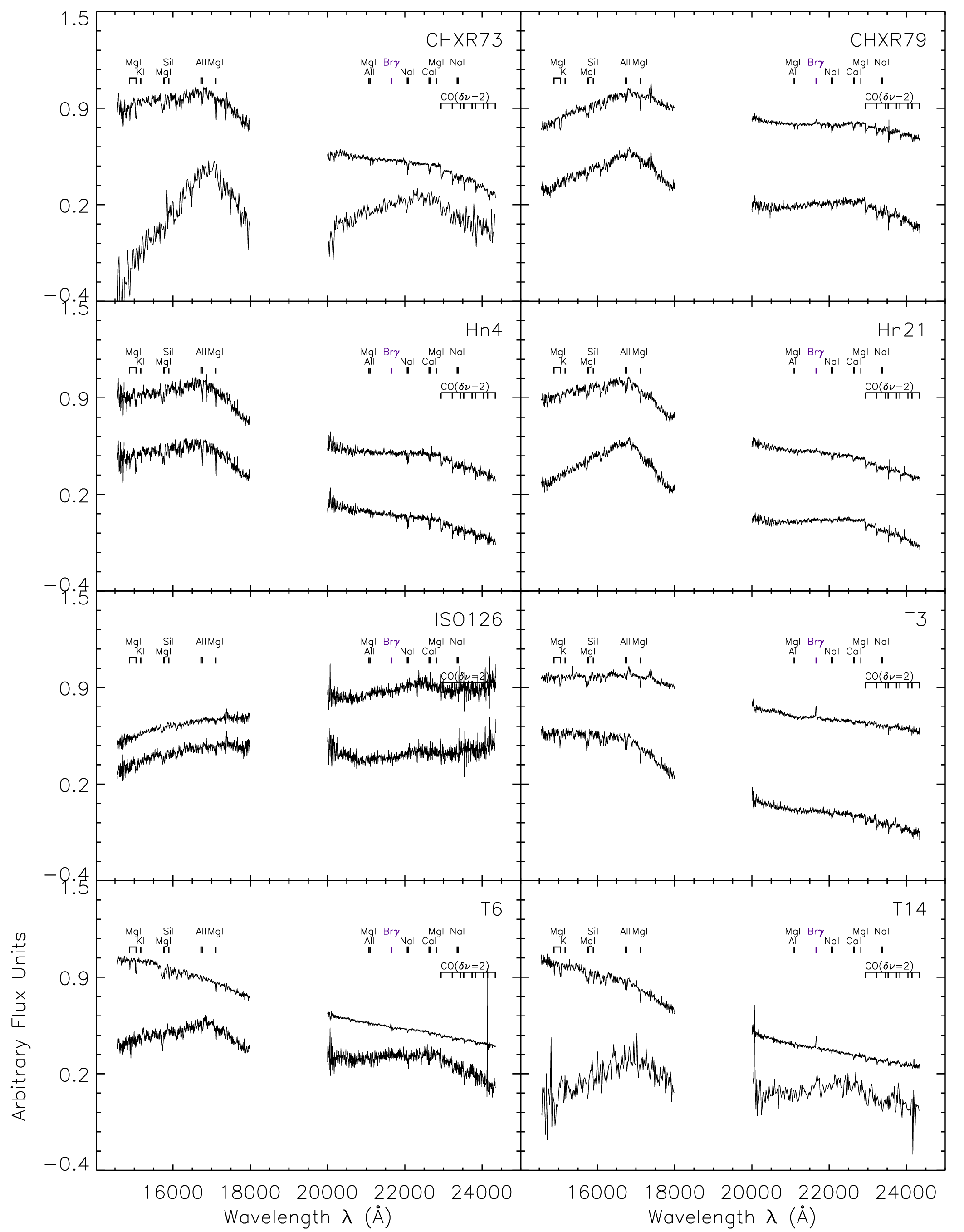}
  \caption[]{ctd.}
\end{figure*}
\addtocounter{figure}{-1}
\begin{figure*}
  \centering
  \includegraphics[angle=0,width=1.0\textwidth]{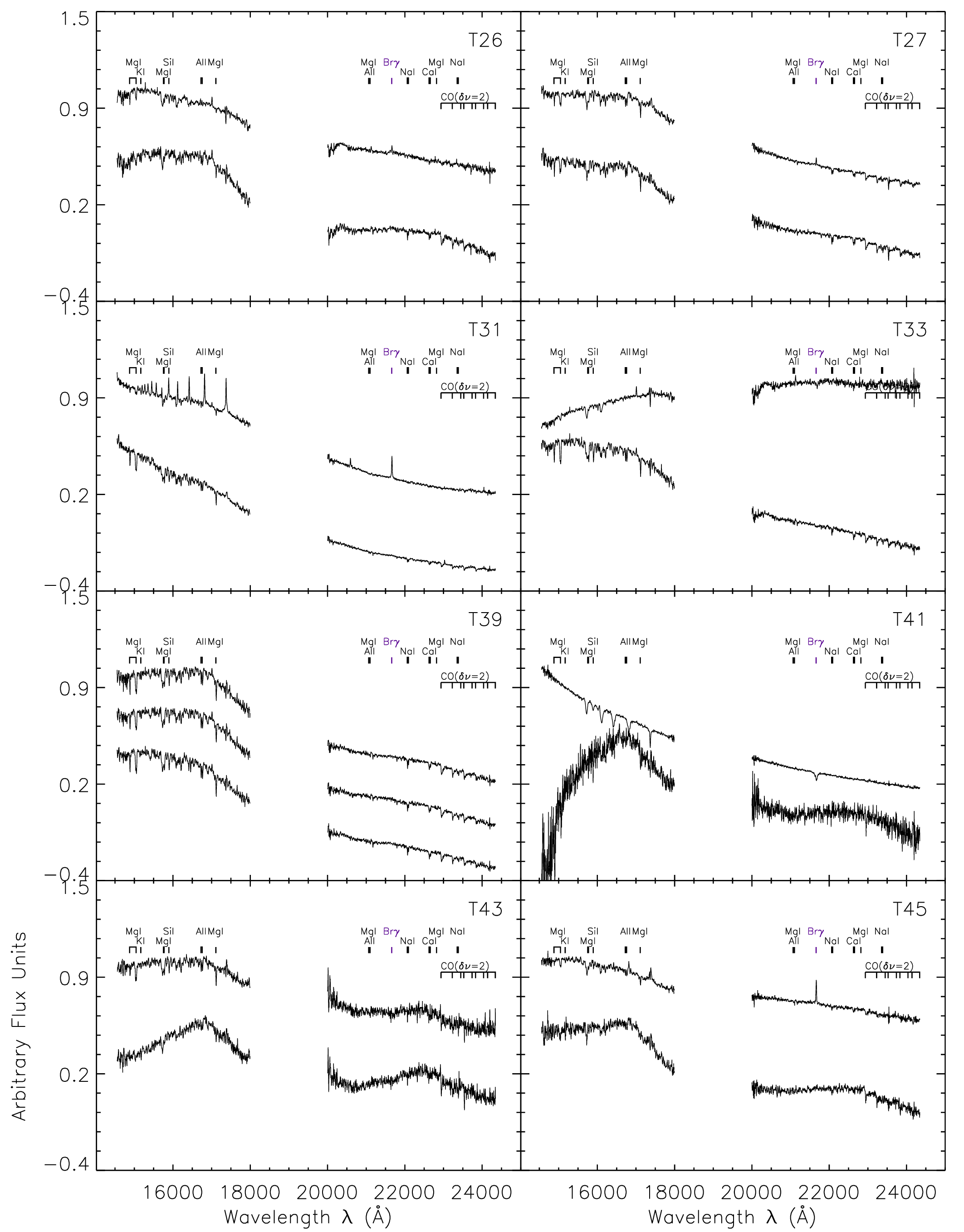}
  \caption[]{ctd.}
\end{figure*}
\addtocounter{figure}{-1}
\begin{figure*}
  \centering
  \includegraphics[angle=0,width=1.0\textwidth]{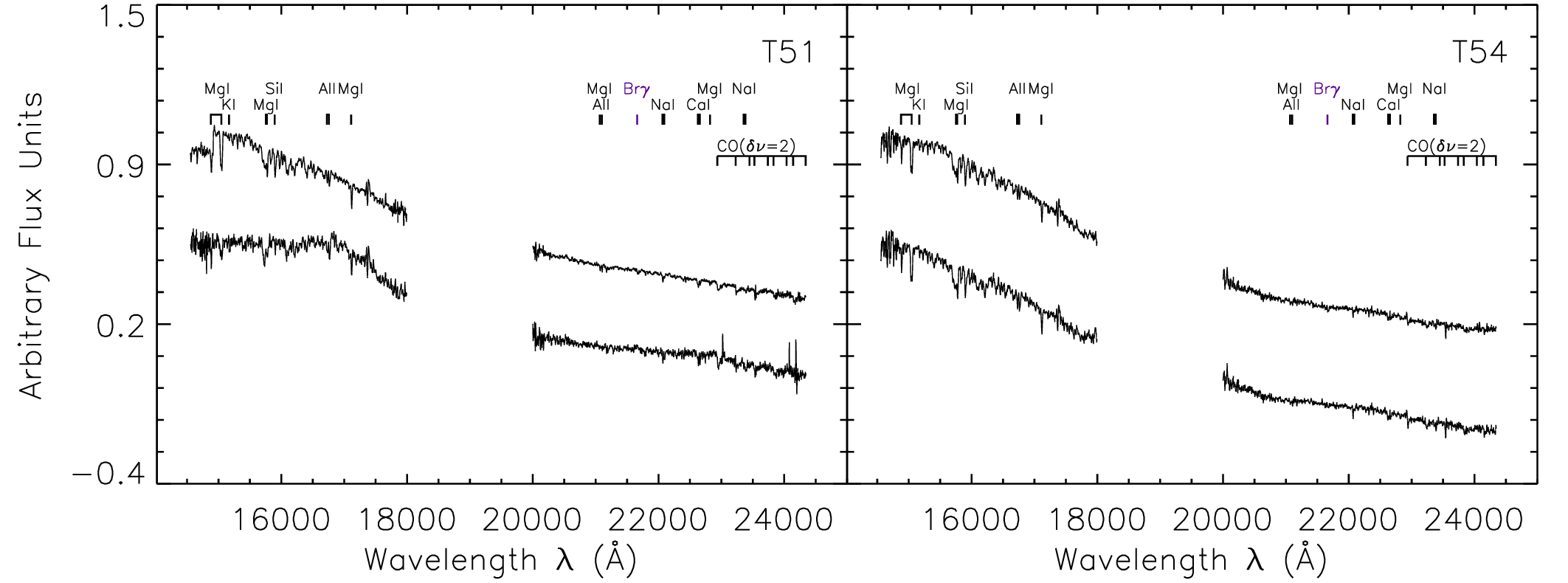}
  \caption[]{ctd.}
\end{figure*}

\end{document}